\documentclass[acmsmall]{acmart}

\AtBeginDocument{%
  \providecommand\BibTeX{{%
    \normalfont B\kern-0.5em{\scshape i\kern-0.25em b}\kern-0.8em\TeX}}}

\setcopyright{acmcopyright}
\copyrightyear{2018}
\acmYear{2018}
\acmDOI{10.1145/3362168}
\acmJournal{JACM}
\acmVolume{37}
\acmNumber{4}
\acmArticle{111}
\acmMonth{8}

\usepackage{multirow}

\begin{document}

\title{Decentralized Trust Management: Risk Analysis and Trust Aggregation}

\author{Xinxin Fan}
\email{fanxinxin@ict.ac.cn}
\orcid{0000-0002-6659-7431}
\affiliation{%
  \institution{Institute of Computing Technology, Chinese Academy of Sciences}
  \streetaddress{No.6 Kexueyuan South Rd}
  \city{Beijing}
  \state{}
  \postcode{100190}
  \country{China}}
\author{Ling Liu}
\email{lingliu@cc.gatech.edu}
\affiliation{%
  \institution{School of Computer Science, Georgia Institute of Technology}
  \streetaddress{266 Ferst Dr}
  \city{Atlanta}
  \state{Georgia}
  \postcode{30332-0765}
  \country{USA}
}

\author{Rui Zhang}
\affiliation{%
  \institution{Institute of Information Engineering, Chinese Academy of Sciences}
  \streetaddress{No.89 Minzhuang Rd}
  \city{Beijing}
  \country{China}}
\email{zhangrui@iie.ac.cn}

\author{Quanliang Jing}
\author{Jingping Bi}
\affiliation{%
  \institution{Institute of Computing Technology, Chinese Academy of Sciences}
  \city{Beijing}
  \state{}
  \postcode{100190}
  \country{China}
}
\email{Jingquanliang@ict.ac.cn, bjp@ict.ac.cn}

\renewcommand{\shortauthors}{Xinxin Fan, et al.}

\begin{abstract}
Decentralized trust management is used as a referral benchmark for assisting decision making by human or intelligence machines in open collaborative systems. During any given period of time, each participant may only interact with a few of other participants. Simply relying on direct trust may frequently resort to random team formation. Thus, trust aggregation becomes critical. It can leverage decentralized trust management to learn about indirect trust of every participant based on past transaction experiences. This paper presents alternative designs of decentralized trust management and their efficiency and robustness from three perspectives. First, we study the risk factors and adverse effects of six common threat models. Second, we review the representative trust aggregation models and trust metrics. Third, we present an in-depth analysis and comparison of these reference trust aggregation methods with respect to effectiveness and robustness. We show our comparative study results through formal analysis and experimental evaluation. This comprehensive study advances the understanding of adverse effects of present and future threats and the robustness of different trust metrics. It may also serve as a guideline for research and development of next generation trust aggregation algorithms and services in the anticipation of risk factors and mischievous threats.
\end{abstract}


\setcopyright{acmcopyright}
\acmJournal{CSUR}
\acmYear{2019} \acmVolume{1} \acmNumber{1} \acmArticle{1} \acmMonth{1} \acmPrice{15.00}\acmDOI{10.1145/3362168}

\maketitle

\section{Introduction}
Trust is an abstract, multi-faceted and subjective concept \cite{136-Liu}. Trust has been investigated in multiple disciplines in addition to computer science, ranging from business, philosophy, social science. Researchers from different domains agree with the fundamental definition of trust, i.e., trust describes an anticipation/trustworthiness level of an individual (as a human being or an intelligence machine). Trust is often derived from certain feedback ratings through trust aggregation. For example, Gambetta \citeyearpar{1-Gambetta} presented trust as the subjective probability in social science that a trustor anticipated a trustee to execute an action beneficial to her/him. Lahno \citeyearpar{2-Lahno} introduced the philosophy of distrust as the betrayal of moral behavior. In economics, trust is reflected by the decision to maximize the trustor's interest by trading off between the potential risks and the possible utility gains \cite{3-Cho}. In computer science, many research branches have adopted trust to mitigate various threats and risks through tracking and leveraging historical interaction experiences in open computing systems and networks. Trust is regarded as an essential pillar for our digital economy and our cyber infrastructure \cite{136-Liu}.

{\bf Trust management} refers to managing trust in a computing system, including defining trust, identifying the elements that establish trust, and mechanisms for trust computation, trust propagation, trust aggregation, trust data storage as well as the usage models of trust and trust enhanced service provisioning. One can provide the above functionalities using a centralized computing architecture or a decentralized computing architecture or a hybrid of centralized and decentralized computing architectures, which allows certain trust functionality to be implemented and supported using distributed computing platforms and distributed computation algorithms. {\bf Decentralized trust management} refers to managing trust in either fully decentralized computing systems or a hybrid of centralized and decentralized computing systems.

Over the last decade, trust management has penetrated diverse collaborative networked computing systems, ranging from peer to peer and eCommence, social networks and online community, cloud and edge computing, mobile ad hoc networks and wireless sensor networks, to crowdsourcing, multi-agent and Internet of things (IoTs) \cite{135-Liu}.

\textbf{Peer-to-Peer Trust.} Trust management in Peer-to-Peer systems has been studied for more than two decades \cite{38-Xiong, 35-Josang, 32-Kamvar, 43-Su, 4-Fan} and surveyed \cite{100-Josang, 105-Suryanarayana} in the context of decentralized overlay networks and applications. Suryanarayana and Taylor \citeyearpar{105-Suryanarayana} compared trust metrics from trust attributes and discovery mechanism. J{\o}sang e al. \citeyearpar{100-Josang} summarized a number of main concerns in establishing and utilizing trust, such as low incentive for providing ratings, biases toward positive feedback, colluding participants, unfair feedback ratings from mischievous participants, changing identities, etc.

\textbf{Multi-Agent Trust.} Trust management in a multi-agent system (MAS) utilizes trust to improve the collaboration among multiple autonomous agents in accomplishing a task. Balaji and Srinivasan \citeyearpar{120-Balaji} defined the trust of an agent in terms of autonomy, inferential capability, responsiveness and social behavior. Granatyr et al. \citeyearpar{121-Granatyr} reviewed trust models for MAS by analyzing a set of trust dimensions, such as trust semantics, trust preference, delegation, risk measure, incentive feedback, initial trust, open environment, hard security threats and requirements, and identified the linking between trust dimensions and types of interactions, such as coalition, argumentation, negotiation and recommendation. Pinyol and Sabater-Mir \citeyearpar{122-Pinyol} reviewed trust metrics in terms of trust cognition, procedure and generality. Yu et al. \citeyearpar{119-Yu} reviewed existing trust metrics from a game theoretic perspective. Braga et al. \citeyearpar{124-Braga} unveiled some characteristics of trust, such as the usage of multiple input sources, cheating assumptions, provision of procedural and cognition concepts.

\textbf{Social Network Trust.} Trust management in social networks and online communities has been an active research in the last decade \cite{20-Caverlee, 117-Wang, 123-Jiang}. Caverlee et al. \citeyearpar{20-Caverlee} proposed social incentives with personalized similarity to improve the aggregation of reputation trust in a large-scale social network in which participants often did not know each other a priori. Sherchan et al. \citeyearpar{61-Sherchan} surveyed several critical attributes of social trust, such as dynamic, propagative, non-transitive properties, interaction behaviors, and historical experiences. Jiang et al. \citeyearpar{116-Jiang} reviewed the graph-based trust evaluation for online social networks (OSNs) in two broad categories: graph simplification-based approaches and graph analogy-based approaches. Jiang et al. \citeyearpar{123-Jiang} also proposed another work to focus on 1-hop recommender selection problems in OSNs, e.g. selecting all/a fixed number of/a fixed proportion of/top \(m\) qualified neighbors.

\textbf{Mobil and Wireless Ad Hoc Network Trust.} Trust is a popular mechanism for secure routing in mobile and wireless ad hoc networks (MANETs). Zhang et al. \citeyearpar{106-Zhang} incorporated trust metrics into the routing protocol in wireless ad hoc networks, and provided a theoretical analysis in the perspectives of correctness, optimality and inter-operativity. Movahedi et al. \citeyearpar{125-Movahedi} reviewed several trust frameworks to tackle the bad-mouthing attack and double-face attack. Several survey articles \cite{126-Cho, 127-Govindan, 128-Tangade, 129-Agrawal} reviewed trust metrics for MANETs and presented a more comprehensive categorization of potential attacks, such as routing loop attack, wormhole attack, blackhole attack, grayhole attack, DoS attack, on-off attack, package modification/insertion, incomplete information, selective misbehaving attack, conflicting behavior attack, etc. These attacks enlarge our horizon on threats and vulnerability risks, and provide the basis for the verification of trust management.

\textbf{Cloud Computing Trust.} Noor et al. \citeyearpar{130-Noor} classified trust management in cloud computing into four categories with respect to the roles of service requester and service provider: (i) policy; (ii) recommendation; (iii) reputation; and (iv) prediction. Ahmed et al. \citeyearpar{131-Ahmed} presented a trust evaluation survey for the cross-cloud federation, namely a federation comprised of unknown cloud service providers with heterogeneous infrastructures sharing resource for a limited period. It argued the overall requirements for trust evaluation should include the special requirement from cross-cloud federation, encompassing the architecture and operational principles of federation.

\textbf{Trust and Cryptography.} Kerrache et al. \citeyearpar{104-Kerrache} proposed an adversary-oriented survey on trust and cryptography for vehicular network in terms of security communication, safety application and infotainment application. The security communication mainly contained certificate replication attack, eavesdropping attack and vehicle/driver privacy attack. The safety application primarily included denial of service (DoS), jamming attack, coalition and platooning attack and betrayal attack. The infotainment application mainly involved replayed/altered/injected message attack and illusion attack, in addition to the common attacks, such as masquerading attack and impersonation attack, Sybil attack, and GPS position faking attack, timing attack and black-hole/gray-hole attacks.

\textbf{Trust from Multi-disciplinary Research.} Trust has been a common theme from a multi-disciplinary perspective. Cho et al. \citeyearpar{3-Cho} surveyed composite trust through deriving trust factors from communication, information, society and cognition, and discussed trust in a comprehensive extent, covering artificial intelligence, human computer interaction, data fusion, human-machine fusion, computer networking and network security, data mining and automation.

{\bf Contributions and Scope.\/} Comparing with existing surveys on trust management in different subject areas above, our paper presents three unique contributions: (1) We provide an in-depth characterization of the inherent vulnerabilities and robustness of existing trust metrics through multi-dimensional analysis and extensive experiments, in addition to the root-cause investigation. (2) We formalize the attack cost and the adverse effect using six representative threat models for risk analysis and trust robustness evaluation with comprehensive experimental verification. (3) By taking into account direct trust aggregation and various trust propagation kernels, we summarize the existing trust metrics into six classifications, and evaluate their robustness and adaptability against the six threat models.

In short, this survey focuses on integrating threat models with trust management and provides an in-depth study of fundamental factors for trust establishment, trust propagation, and trust utility in the presence of six categories of common risks and threats. It can serve as a guideline for research and development of next generation trust aggregation algorithms and assist human or intelligence machines to leverage trust management for making decisions in the anticipation of various risk factors and mischievous threats.

First, we introduce how to establish direct trust among individual participants in an interactive network. We review different methods to derive indirect trust from direct trust information, and introduce common reference model for trust establishment, such as honest/dishonest rating, non-creditable rating, feedback credibility-weighted (FCW) direct trust, uniformly distributed trust propagation and threshold-controlled trust propagation kernels.

Second, we categorize common threats and risks emerged in trust management for diverse interactive networks into six types of threat models to characterize six differential types of mischievous adversaries. We quantitatively infer the adverse effect and attack cost for each threat model with experimental analysis and demonstration.

Third, we provide an in-depth analytical comparison of the state of the art trust research from two core-components of decentralized trust management systems: direct trust aggregation and trust propagation kernel design. We study the inherent vulnerabilities of existing trust metrics and evaluate their attack resilience in the context of the six threat models. In addition, we summarize existing trust metrics into six categories based on different trust propagation kernels and direct trust aggregation fashions.

{\bf Organization.\/} The rest of this survey paper is organized as follows. We first describe and compare the state of the art research in trust management from direct trust aggregation and indirect trust aggregation in Section \ref{sec:two} and Section \ref{sec:three} respectively. Then we describe and categorize attacks and risks into six threat models and present quantitative analysis of the adverse effect and attack cost for the six threat models in Section \ref{sec:four}. We compare different trust aggregation kernels and study the root-causes of their vulnerabilities in the presence of the six threat models in Section \ref{sec:five} and provide design principle of decentralized trust management in Section \ref{sec:six}. We describe trust applications in edge computing, blockchain and trust data storage in Section \ref{sec:seven}, and conclude this survey in Section \ref{sec:eight}.
\section{DIRECT TRUST ESTABLISHMENT} \label{sec:two}
\subsection{Transactions and Interactions in Distributed Open Systems}
We can broadly classify distributed networked systems into four categories: (1) distributed clients and centralized servers; (2) distributed clients and distributed servers, and no direct communications between distributed clients;  (3) the hybrid, supporting client-server communication in both peer to peer among distributed clients in addition to those from clients to centralized severs or distributed servers; and (4) the peer to peer decentralized networked system, where no centralized servers are supported and all client-server communications are done among distributed nodes that serve dual roles of client and server. Web service provisioning from enterprises, such as Amazon, Uber, Airbnb, would belong to the distributed networked systems of type (1) or type (2), and the systems of type (1) and type (2) offer centralized services to a large population of distributed clients. Skype and WeChat are examples of type (3) and the systems of type (3) provide both centralized servers and decentralized servers as their service provisioning platforms. Bitcoin and Tor are representative peer to peer systems of type (4) and the systems of type (4) have no centralized management and participants of the type (4) systems form a peer to peer overlay network with decentralized routing protocols (e.g., such as neighbor based broadcast) to reach the rest of the network and share resources with, or provide services to, the other participants in the network.

In this paper we primarily focus on decentralized trust management in distributed networked systems of type (3) and type (4). By decentralized trust management, we mean that the majority of the trust management functionalities, such as trust establishment, trust computation, trust aggregation, trust propagation, will be provided using a decentralized computing architecture. For instance, a participant in an open network can issue a query service to the network as a client to search for a resource, some other participants in the network may respond and provide the resource as the server. Considering the dynamics of open networks, participants may join or depart the network randomly, and the network structure is continuously changing over time. Thus, different requests may require different sets of participants to work together for effective service provisioning. Assuming each request is being served by one sever even when multiple participants may be able to provide the same service. We refer to each service interaction with the completion of a service request as a transaction query between a pair of participants, one as a client and the other as a server. Once one transaction is accomplished, the client participant may give its feedback rating on the server participant with respect to the quality of the transaction query \(Q\). Within open networks, each participant can be a client or a server to a \(Q\) and also can be a feedback rater if it is a client participant for \(Q\) or a feedback rating receiver if it is a server participant.

Trust management in an open system can also be categorized into three categories based on the three types of interaction patterns: Human to Human (H2H), Human to Machine (H2M) or Machine to Machine (M2M).

\textbf{Human-to-Human (H2H).} H2H trust represents the trustworthy relationship among humans in a physical or virtual community, in which individuals employ a computer-assisted networked system to establish interaction among or with friends and families, as well as unknown individuals in social, physical or virtual community. The direct trust for a pair of participants in such a community reflects the actual interactions driven by common social or business interest, direct or indirect friendship through shared background or experiences, etc. It heavily reflects the human attributes, e.g., emotion, intimacy and mutual reciprocity~\cite{60-Granovetter}. H2H trust management illuminates the complex trustworthiness relationships in a variety of disciplines, including anthropology~\cite{61-Sherchan}, sociology \cite{62-Golbeck, 63-Golbeck, 64-Golbeck}, economics \cite{65-Xiong, 66-Zhang}, social psychology~\cite{67-Rotter, 68-Cook} and organizational studies \cite{70-Jackson}.

\textbf{Human-to-Machine (H2M).} H2M trust describes the trustworthiness between humans and machine hosted services in a computer-aided networked system. H2M trust management can be regarded as an interface between human consumers and machine-supported services, which on one hand assists consumers to select trustworthy services, and on the other hand, prevents consumers from getting untrusted services or blocks attacks to the service hosting system. Noor et al. \citeyearpar{77-Noor} proposed to aggregate the consumers' feedback ratings on cloud services. Habib et al. \citeyearpar{78-Habib} proposed a trust mechanism to guarantee clients to receive only trustworthy cloud services. Hwang and Li \citeyearpar{79-Hwang} built a trust-based cloud service architecture to protect both cloud providers and consumers. Li et al. \citeyearpar{80-Li} proposed a trust-aware service brokering system to assist the selection of trustworthy cloud services.

\textbf{Machine-to-Machine (M2M).} M2M trust management refers to the mechanisms that measure and manage the trustworthiness of the functionalities performed by the participants of the open networked system, which are typically software agents hosted on the network nodes (virtual or physical machines). When the pair of communicating nodes can accomplish the intended transaction with an expected result \cite{135-Liu}, the client machine can provide a good feedback rating for the server machine based on the expected behavior, typically defined by the trust model in the form of trust policies. Walter et al. \citeyearpar{87-Walter} proposed a trust-based recommendation approach to assist the selection of well-behaved vehicles on demand. Tan et al. \citeyearpar{88-Tan} proposed a trust management scheme to secure the data plane of ad-hoc networks. Nitti et al. \citeyearpar{94-Nitti} proposed to use trust metrics to assist the establishment of trustworthy IoTs.

In this paper, although we focus primarily on decentralized trust management in machine to machine communication scenarios, many of the design principles and trust management algorithms such as trust aggregation and trust propagation kernels can be easily adapted to the H2M and H2H trust management systems.
\subsection{Direct Trust with Local Trust Aggregation}
Generally, the feedback rating is positive if the transaction query is satisfied, or negative if unsatisfied. Nevertheless, the emergence of strategically mischievous participants (SMPs) breaks this routine feedback pattern, i.e. the SMPs on the one hand provide high-quality transaction queries to get honest (positive) ratings from other service receivers as server participants, but oppositely give dishonest (negative) feedback ratings to other service providers as client participants ignoring whatever the transaction queries are satisfied or unsatisfied \cite{4-Fan, 32-Kamvar, 33-Su, 34-fan}. In addition, the query transaction may fail with non-response, or delivering faulty/low-quality results due to unintended reasons, such as network bandwidth jitter, cooling-induced cloud server downtime, etc. Upon the above analysis, we know that the mischievous behavior can be studied from i) two-facet intended manners, i.e. service-based misbehaved manipulation, rating-based misbehaved manipulation; and ii) one unintended manner, i.e. system/network reliability factors-induced natural failure.

\textbf{Service-based Feedback Rating.} Normally, we refer the feedback that a service provider receives a positive/negative rating while providing an authentic/inauthentic service as honest rating. Inversely, we refer the feedback that a service provider receives a negative/positive rating while providing an authentic/inauthentic service as dishonest rating. Each individual can alternately play the two roles during the interactions, i.e. service provider (a.k.a. server participant) and service consumer (a.k.a. client participant). The vicious participants can be categorized as independently mischievous, collusively mischievous, randomly mischievous, occasionally mischievous and persistently mischievous \cite{4-Fan}. In fact, the rating is reflected by feedback rater being honest or dishonest, being independently or collusively dishonest, being randomly or occasionally dishonest or persistently dishonest, but all about the rater's perception on the query transaction quality.

The different categories of mischievous participants may have some overlap in terms of malicious manipulation, such as they all provide inauthentic services and dishonest feedback ratings, but for different threat models the malicious behaviors may be naively malicious or strategically malicious in serving or rating or both. Prior to establishing direct trust, we first give some basic definitions.

\begin{definition}[Honest Rating] The rating is strictly subject to the query transaction quality, i.e. positive rating for authentic service and negative rating for inauthentic service. It can be trusted by the network as a local trust metric for the feedback receiver.
\end{definition}

\begin{definition}[Dishonest Rating] The rating is alternatively subject to the transaction target rather than the truthful query transaction quality, i.e. positive rating for colluding participant and negative rating for routine participant. It ought to be weighted by feedback credibility prior to being trusted by the network.
\end{definition}

Usually, 5\% dishonest ratings are allowed to reflect the randomly or occasionally dishonest behavior of an honest rater \cite{4-Fan, 32-Kamvar} due to some unintended reasons. The direct trust from a participant \(p_i\) to another participant \(p_j\) based on one-time transaction can be demonstrated using binary or multiscale rating. For example, the pioneering trust metric EigenTrust \cite{32-Kamvar} defined the direct trust for a pair of transacted participants using binary rating \{-1, +1\}, \(tr(p_i, p_j)\)=-1 denoted a negative rating from \(p_i\) to \(p_j\), and \(tr(p_i, p_j)\)=+1 represented a positive rating. The heritage EigenTrust\(^{++}\) \cite{34-fan} and GroupTrust \cite{4-Fan}, both employed this kind of binary rating. Differently, ServiceTrust \cite{33-Su} and ServiceTrust\(^{++}\) \cite{43-Su} utilized the multiscale rating \{-1, 0, 1, 2, 3, 4, 5\} indicating bad, no-rating, neutral, fair, good, very good and excellent query transaction respectively.

For SMPs, one smart way to subvert the system is to firstly collect positive ratings to yield high trust through providing authentic services, then utilize the advantage of gained trust to participate the query transactions to provide inauthentic services. Hence, a reliable trust metric ought to take the historical feedback information into account, i.e. recent feedback ratings should be more valuable and historical feedback ratings should be less valuable. The studies \cite{95-Li, 96-Li, 90-Wang, 91-GiorgosZacharia} also confirmed the ratings in a recent time period weighted more than the former emerged ratings. In this way, once SMPs are found to change their transactional behaviors from honestly providing authentic services to dishonestly offering inauthentic services, their trust would be degraded shortly using weight parameter even they possessed high trust already. For instance, Li et al. \citeyearpar{9-Li} used the weight decimal [0, 1] manner to distinguish the different-time feedback ratings for personal rating and indirect recommendation. For a given period of time interval [\(t_1\), \(t_n\)], the direct trust level from \(p_i\) to \(p_j\) can be calculated as:
\begin{equation}
tr_h^{} (p_i ,p_j )^{(t_i )} = w_{t_i } \cdot tr(p_i, p_j )^{(t_i )} \quad  t_i \in [t_1 ,t_n],
\end{equation}
where \(w_{t_i }\) set as \(a^{t_n  - t_i}\) (\(0 < a \le 1\)) is the weight of the rating \(tr(p_i ,p_j )^{(t_i )}\) at time \(t_i\).

\textbf{Rating-based Feedback Rating.} The SMPs can indeed earn high trust by aggregating honest positive ratings from good participants via serving authentic resources and dishonest positive rating from their colluders. This violates the initial aim of trust metrics at degrading the trust levels of mischievous participants. Thus only utilizing honest rating is inadequate for the SMPs, we need further explore creditable rating to amend the deficiency.

\begin{definition}[Creditable Rating] The rating for a single feedback rater or pairwise feedback rating score is integrated by feedback credibility factor.
\label{definition-3}
\end{definition}

Oppositely, non-creditable rating can be defined as:

\begin{definition} [Non-creditable Rating] The rating for a single feedback rater or pairwise feedback rating score is always negative without referring to any creditable factor.
\end{definition}

Upon Definition \ref{definition-3}, we can further study the creditable ratings from two levels, namely feedback rater and feedback rating score.

\begin{definition} [Feedback Rater Level Credibility] The interactive system endows each feedback rater a credibility weight for local trust aggregation.
\end{definition}

\begin{definition} [Feedback Rating Score Level Credibility] The interactive system endows each feedback rating score over a pair of transacted participants a credibility weight for local trust aggregation.
\end{definition}

Xiong and Liu \citeyearpar{38-Xiong} had proposed two kinds of credibility measure fashions from the standpoints of feedback rater level and feedback rating score level. The former (PeerTrustTVM) set self-trust as credibility weight, which interpreted the feedback rating of a trustworthy participant possessed more credibility than that of an untrustworthy participant.
\begin{equation}
Cr_{p_i } = \frac{{T(p_i )}}{{\sum\nolimits_{p_m  = 1}^{|tr(p_j )|} {T(p_m )} }} \quad p_i \in tr(p_j ),
\end{equation}
where \(tr(p_j)\) represented the set of participants that had transactions with \(p_j\), \(T(p_i)\) denoted the trust value of \(p_i\). The latter (PeerTrustPSM) employed feedback similarity as credibility factor, the mischievous participants had low feedback similarity with good participants due to greatly different ratings to commonly transacted participants.
\begin{equation}
Cr_{p_i p_k }  = \frac{{sim(p_i ,p_k )}}{{\sum\nolimits_{p_m  = 1}^{|tr(p_j )|} {sim(p_i ,p_m )} }} \quad p_i ,p_k  \in tr(p_j ),
\end{equation}
\begin{equation}
sim(p_v ,p_w ) = 1 - \left( {\frac{{\sum\limits_{p_x  \in comn(p_v ,p_w )} {(\sum\limits_{|tr(p_v ,p_x )|} {\frac{{tr(p_v ,p_x )}}{{|tr(p_v ,p_x )|}}}  - \sum\limits_{|tr(p_w ,p_x )|} {\frac{{tr(p_w ,p_x )}}{{|tr(p_w ,p_x )|}}} )^2 } }}{{|comn(p_v ,p_w )|}}} \right)^{\frac{1}{2}},
\end{equation}
where \(sim(p_v, p_w)\) was the feedback similarity through inferring the standard deviation of feedback ratings to the commonly rated participants \(comn(p_v, p_w)\). Similarly, GroupTrust \cite{4-Fan} used exponential function to define feedback rating score level credibility:
\begin{equation}
Cr'_{p_i p_j }  = exp\{ 1 - \frac{1}{{sim(p_i ,p_j )}}\}.
\end{equation}

\textbf{Raw Local Trust Aggregation.} The direct trust over each pair of transacted participants can be calculated via local trust aggregation. At present, the commonly used aggregating fashions can be roughly classified into two manners: i) transaction success ratio; and ii) beta function probability expectation. Intuitively, the transaction success ratio-employed direct trust from \(p_i\) to \(p_j\) can be defined as:
\begin{equation}
s_{p_i p_j }  = \left\{ \begin{array}{l}
 \frac{{\delta _{p_i p_j } }}{{\delta _{p_i p_j }  + \sigma _{p_i p_j }  + 1}}_{} \frac{{\sigma _{p_i p_j } }}{{\delta _{p_i p_j }  + \sigma _{p_i p_j }  + 1}} \le \theta  \\
 \begin{array}{*{20}c}
   {} & {}  \\
\end{array}\begin{array}{*{20}c}
   {\frac{1}{2}} & {}  \\
\end{array}_{} otherwise \\
 \end{array} \right.,
\end{equation}
where \(\theta\) implies good participants misbehave in a tiny probability due to system reliability factors-induced natural failure, usually set as 5\% \cite{32-Kamvar, 33-Su, 34-fan}. \(\delta _{p_i p_j}\) denotes the number of successful transactions between \(p_i\) and \(p_j\), \(\sigma _{p_i p_j }\) is the number of unsuccessful transactions.

The beta probability density functions \cite{35-Josang, 59-Klos, 87-Walter} can be expressed via gamma function:
\begin{equation}
f(p|\alpha ,\beta ) = \frac{{\Gamma (\alpha  + \beta )}}{{\Gamma (\alpha )\Gamma (\beta )}}\rho ^{\alpha  - 1} (1 - \rho )^{\beta  - 1},
\end{equation}
where \(\alpha\), \(\beta\)\(>\)0, and 0 \(\le\) \(\rho\) \(\le\) 1, \(\rho\) \(\ne\) 0 if \(\alpha\) \(<\) 1 and \(\rho\) \(\ne\) 1 if \(\beta\) \(<\) 1. At present, scholars straightly define the direct trust as the probability expectation of beta distribution:
\begin{equation}
\label{equation-8}
s_{p_i p_j }^\beta   = \alpha /(\alpha  + \beta ) = (\delta _{p_i p_j }  + 1)/(\delta _{p_i p_j }  + \sigma _{p_i p_j }  + 2),
\end{equation}
where \(\alpha\) = \(\delta _{p_i p_j }\)+1, \(\beta\) = \(\sigma _{p_i p_j }\)+1. Essentially, this beta function based direct trust reflects the transaction success ratio as well. For easy understanding, we call both \(s_{p_i p_j}\) and \(s_{p_i p_j }^\beta\) as raw direct trust with different local trust aggregation fashions given they only adopt positive and negative ratings to produce direct trust for a pair of transacted participants.

Without a doubt, the local trust aggregation can pull-in some more crucial impact factors, e.g. the aforementioned history feedback factor and feedback credibility. Naturally, referring to historical feedback factor, the local trust aggregation \(s_{p_i p_j }^h\) can be defined as:
\begin{equation}
s_{p_i p_j }^h  = \frac{{\sum\nolimits_{i = 1}^n {tr_h (p_i ,p_j )^{(t_i )} } }}{{\sum\nolimits_{i = 1}^n {w_{t_i } } }} \quad t_i \in [t_1 ,t_n].
\end{equation}

Furthermore, referring to feedback credibility factor we can define the FCW direct trust.

\begin{definition} [Feedback Credibility-Weighted Direct Trust] The direct trust over each pair of transacted participants is yielded through integrating creditable trust factors from the viewpoint of a single feedback rater or pairwise feedback rating score.
\end{definition}

Accordingly, the feedback rater level credibility-based direct trust \({s'}_{p_i p_j }^{Cr}\) is defined as:
\begin{equation}
{s'}_{p_i p_j }^{Cr}  = Cr_{p_i }  \cdot s_{p_i p_j } \quad p_i \in tr(p_j).
\end{equation}

The pairwise feedback rating score level credibility-based direct trust \(s_{p_i p_j }^{Cr}\) (\(s_{p_i p_j }^{Cr'}\)) is defined using feedback similarity among the two transacted participants \cite{4-Fan}:
\begin{equation}
s_{p_i p_j }^{Cr'}  = Cr'_{p_i p_j }  \cdot s_{p_i p_j } \quad p_i, p_j \in tr(p_k),
\end{equation}
or using feedback similarity of reference third-party participants \cite{38-Xiong}:
\begin{equation}
s_{p_i p_j }^{Cr}  = Cr_{p_i p_k }  \cdot s_{p_k p_j } \quad p_k, p_i \in tr(p_j).
\end{equation}

Trust itself is a complex and subjective concept impacted by multiple factors with respect to the diversities of misbehaved participants, thereby, for mischievous participants especially SMPs, it is hard to constrain them from gaining high direct trust from honestly transacted participants. Therefore, an effective trust metric need take into account these impact factors to infer a multi-perspective and rational direct trust for each two transacted participants.
\section{INDIRECT TRUST WITH NETWORK-BASED TRUST AGGREGATION} \label{sec:three}
\subsection{Network-Scoped Trust Aggregation}
The baseline inference on indirect trust from participants \(p_u\) to \(p_v\) is to recursively aggregate the third-party participant \(p_k\)'s direct trust placed on \(p_v\) within the holistic network, i.e. \(\sum\limits_{p_k } {s_{p_u p_k } \cdot s_{p_k p_v } }\) or \(\sum\limits_{p_k } {s_{p_u p_k }^{Cr} \cdot s_{p_k p_v }^{Cr} }\). For an interactive network with \(N_{net}\) participants, the network-scoped trust can be derived using the power iteration of adjacent matrix in which each element stands for the direct trust value over each pair of participants.
\begin{equation}
T_G ^{(k + 1)}  = M^T  \cdot T_G ^{(k)},
\end{equation}
where \(T_G^{(k+1)}\) denotes the (\(k\)+1)th iteration trust vector of \(N_{net}\) participants, \(M\) is the normalized direct trust matrix: \(m_{p_u p_v }\)=\(s_{p_u p_v } /\sum\nolimits_{p_m } {s_{p_u p_m } }\), if \(\sum\nolimits_{p_m } {s_{p_u p_m } } \ne\)0, otherwise \(m_{p_u p_v }\)=0. This iteration operation in fact is also a trust propagation/diffusion process hop by hop, \(k\) controls the propagating scope of trust. Obviously, each participant's trust can be propagated to the whole network with certain iteration rounds. Inversely, each participant can also receive trust from the whole network. Thus, we define the global trust as follows.

\begin{definition} [Global Trust] We define the trust score computed by indirect trust aggregation over the entire network through the network topology as the global trust from one participant $p_u$ to another participant $p_v$, provided that $p_v$ is reachable from $p_u$ by network traversal. The global trust value can be viewed as the comprehensive confidence that the entire network as a community places on the participant $p_v$ via the view of $p_u$.
\end{definition}

EigenTrust \cite{32-Kamvar} is the first to introduce the use of the pre-trusted nodes as the authority participants in order to address the "cold start" problem. Taking into account the pre-trusted participants, the authors defined the eigenvector-based global trust as:
\begin{equation} \label{equation-14}
T_G ^{(k + 1)}  = (1 - \varepsilon) \cdot M^T  \cdot T_G ^{(k)}  + \varepsilon  \cdot \overrightarrow P,
\end{equation}
where \(\varepsilon\) denoted the probability a stranger or newcomer would like to trust the system-generated pre-trusted participants \(P\), \(p_{p_j}\)=1/\(|P|\) if participant \(p_j\)\(\in\)\(P\), otherwise \(p_{p_j}\)=0.
\subsection{Trust Propagation Kernel}
From Formula (\ref{equation-14}), we know that an individual's global trust is aggregated through asking the other participants' feedback ratings placed on this individual. We define this kind of trust propagation kernel as uniformly distributed trust propagation.

\begin{definition} [Uniformly Distributed Trust Propagation, UDTP] For each participant, it propagates self-global trust to all the neighboring participants in the light of direct trust values placed on the neighboring participants.
\end{definition}

Although this UDTP kernel is the core to aggregate global trust at present, it confronts rigorous inherent vulnerabilities. It is this UDTP kernel that enhances the global trust scores of SMPs, i.e. they can repeatedly gain high trust deriving from authentic-service provision activities. If no mischievous participant exists, this UDTP kernel can yield correct and rational trust level for each participant. To address the vulnerability, a threshold-controlled trust propagation kernel was proposed in \cite{4-Fan, 43-Su}.

\begin{definition} [Threshold-Controlled Trust Propagation, TCTP] For a participant, the decision whether it can propagate trust to its neighboring participants strictly depends on the system-inferred critical threshold.
\end{definition}

Generally, TCTP kernel adopts the direct trust to compare with the system-inferred critical threshold \cite{4-Fan, 43-Su}, if larger, trust propagation is permitted, otherwise trust propagation is blocked. Appropriately setting on TCTP can validly control trust propagation; otherwise it puts TCTP at a disadvantage, i.e. if the threshold is too low, it cannot block SMPs receiving trust propagation from good participants; if the threshold is too high, it might block trust propagation among good participants. An attack resilient trust metric should have the capability to differentially propagate trust among good and different categories of mischievous participants rather than simply utilizing UDTP kernel.
\section{THREAT MODELS AND ADVERSE EFFECTS} \label{sec:four}
\subsection{Reference Threat Models}
Massive simple/strategic threats and risks have been penetrating open networked systems, such as bad-mouthing \cite{31-Sun}, self-promoting \cite{101-Hoffman}, ballot stuffing behavior \cite{29-He, 37-Hu}, collusively malicious \cite{4-Fan, 31-Sun, 40-Jiang}, on-off attack \cite{8-Chae, 31-Sun}, Sybil \cite{6-Wang, 7-Liu}, spy \cite{4-Fan, 32-Kamvar}, black-hole and grey-hole attacks \cite{104-Kerrache}, etc. For the sake of easy understanding, according to references \cite{4-Fan, 32-Kamvar, 43-Su} we summarize these threats and risks as several threat models referring to attack policies and characteristics. Besides the four representative threat models, we also in advance propose another two more-sophisticated threat models to support our deep arguments.

\begin{definition} [Threat Model A-Independently Mischievous] All mischievous participants perform bad services and dishonest feedbacks independently. Concretely, they provide inauthentic services when selected as transaction service providers (server participants) and  they always offer non-creditable ratings to other transacted participants ignoring whether the received services are authentic or inauthentic as feedback raters. The mischievous participants in this category always receive bad ratings from good participants.
\end{definition}

\begin{definition} [Threat Model B-Collectively Mischievous] All mischievous participants are organized in a chain to collude with each other. They always give inauthentic services as server participants, and they always provide dishonest ratings as feedback raters, i.e., giving dishonest (negative) ratings to good participants but dishonest (positive) ratings to other colluding participants over the chain. This Threat Model B adds colluding effect on feedback ratings compared to the Threat Model A.
\end{definition}

\begin{definition} [Threat Model C-Camouflage Collective] All mischievous participants are organized in a chain to collude mutually. They give authentic services in a probability \(f\) when selected as server participants, and provide dishonest ratings as feedback raters, i.e. giving dishonest (negative) ratings to good participants but dishonest (positive) ratings to colluding participants in the chain. The Threat Model C adds {\em a camouflage strategy}: Instead of providing bad services all the time, mischievous participants play camouflage games at a probability \(f\), aiming to cheat the trust system, i.e., the mischievous camouflage participants will receive honest (positive or negative) ratings from some good participants and thus gain relatively higher trust through aggregating positive ratings.
\end{definition}

\begin{definition} [Threat Model D-Group-based Spies] All mischievous participants are divided into two types: Type B acting like the vicious participants in Threat Model A (provide bad services and give dishonest feedbacks) and Type D vicious participants do good services but give dishonest feedbacks. It changes the malicious method of selectively providing good services in Threat Model C to use a subset of mischievous participants to provide good service all the time, but give dishonest feedbacks. It is another malicious strategy intended to cheat the trust system. This threat model adds {\em another malicious strategy}: Type-D participants give negative ratings to good participants but positive ratings to all Type-B participants.
\end{definition}

\begin{definition} [Threat Model E-Camouflage Collective with Honest Rating] Alike Threat Model C, all mischievous participants are organized in a chain to collude each other, but they play camouflage game at both service provisioning and feedback rating. They give authentic services in a probability \(f\) when selected as server participants and will offer honest ratings with probability \(\eta\) as feedback raters. This is a change to Threat Model C with the goal of providing a {\em third malicious strategy} to cheat the trust system: By playing camouflage game also as feedback raters, it allows the good participants to receive positive ratings from camouflage participants at probability \(\eta\), making it hard for the trust system to detect and identify mischievous participants.
\end{definition}

\begin{definition} [Threat Model F-Group-based Spies with Honest Rating] All mischievous participants are composed of Type-D participants and Type-B participants. When selected as server participants, Type-D participants always give authentic services and Type-B participants always give inauthentic services. On the other hand as feedback raters, the mischievous participants offer honest ratings with probability \(\gamma\), that is to say the good participants can receive positive ratings from mischievous spies with probability \(\gamma\). This is a change to Threat Model D with the goal of providing the {\em fourth malicious strategy} to cheat the trust system: The good participants can receive positive ratings from spy participants at probability \(\gamma\), making it hard for the trust system to detect and identify mischievous participants.
\end{definition}

Threat Models A-D have been used by some existing trust metrics, e.g. EigenTrust \cite{32-Kamvar}, ServiceTrust \cite{33-Su}, etc. Threat Models E and F are introduced in this survey to support the most strategically malicious attacks. Threat Models A and B are simple, Threat Models C and D are somewhat more sophisticated, and Threat Models E and F are most strategically malicious, as shown in Table \ref{table:two}.
\subsection{Adverse Effect and Cost of Attacks}
\subsubsection{Attack Cost and Attack Success Ratio}
To formally infer adverse effect and attack cost for each threat model, we first give the definition of attack cost with respect to interactive properties.

\begin{definition} [Attack Cost] Attack cost is comprehensively reflected by the price the mischievous participants need to pay for launching an attack successfully in terms of the amounts of mischievous participants and dishonest feedback ratings, the numbers of authentic services and honest ratings offered by SMPs.
\end{definition}

For each participant, its global trust in fact contains both trustworthiness and untrustworthiness factors, this is because the direst trust over each pair of transacted participants as the base of global trust calculation, is inferred using both positive and negative ratings. To interpret adverse effect, we need separately derive the inherent trust ingredient contribution and distrust ingredient contribution.

\begin{definition} [Trust Ingredient] For a participant \(p_i\), its trust ingredient \(T_{ti}(p_i)\) is aggregated by the honest feedback ratings it received.
\end{definition}
\begin{equation}
T_{ti} (p_i ) = \sum\nolimits_{tr(p_w ,p_i ) \in R_H^{} } {m_{p_w p_i }  \cdot T(p_w )},
\end{equation}
where \(R_H\) is the set of positive feedback ratings offered by good participants or SMPs.
\begin{definition} [Distrust Ingredient] For a participant \(p_i\), its distrust ingredient \(T_{di}(p_i)\) is aggregated by the dishonest and non-creditable feedback ratings it received.
\end{definition}
\begin{equation}
T_{di} (p_i ) = \sum\nolimits_{tr(p_u ,p_i ) \in R_{dH}^{}  \cup R_{nH} } {m_{p_u p_i }  \cdot T(p_u )}
\end{equation}
where \(R_{dH}\) denotes the set of negative ratings offered by collectively mischievous, camouflage and spy participants in Threat Models B-F, \(R_{nH}\) represents the set of non-creditable ratings offered by independently mischievous participants in Threat Model A.

Next, we utilize trust ingredient and distrust ingredient to define attack success ratio.

\begin{definition} [Attack Success Ratio] Given an attack target participant \(p_i\), once the distrust ingredient it received is larger than the trust ingredient, that is to say this participant is successfully attacked. Accordingly, the attack success ratio \(As(p_i)\) is defined as:
\end{definition}
\begin{equation} \label{equation-17}
As(p_i ) = T_{di} (p_i )/T_{ti} (p_i ).
\end{equation}

Obviously, if the adversary participants want to launch an attack successfully on a target \(p_i\), they must yield larger distrust ingredient compared to trust ingredient, i.e. \(As(p_i)\)\(>\)1. For clear description, we show the related notations and presentations in Table \ref{table:one}.
\begin{table}%
\caption{Notations and presentations}
\label{table:one}
\begin{minipage}{\columnwidth}
\begin{center}
\begin{tabular}{lll}
  \toprule
  \multicolumn{1}{c}{Notation}             & \multicolumn{1}{c}{Presentation}   \\
  \hline
  \(N_{net}\)          & number of system participants (network size) \\
  \(R_{dH}\)           & set of dishonest ratings offered by collective, camouflage and spy attackers \\
  \(R_{nH}\)           & set of non-creditable ratings offered by independent attackers \\
  \(R_H\)              & set of honest rating offered by good participants and SMPs  \\
  \(T_G(p_i)\)         & global trust score of participant \(p_i\) \\
  \(T_{ti}(p_i)\)      & trust ingredient of participant \(p_i\)  \\
  \(T_{di}(p_i)\)      & distrust ingredient of participant \(p_i\)  \\
  \(T_{net}^G\)        & network level average trust of good participants \\
  \(T_{net}^M\)        & network level average trust of mischievous participants \\
  \(N_{dH}\)           & number of participants that offer dishonest ratings \\
  \(N_{nH}\)           & number of participants that offer non-creditable ratings \\
  \(N_H\)              & number of participants that offer honest ratings \\
  \(N_C\)              & number of camouflage participants launched by Threat Models C and E \\
  \(N_D\)              & number of Type-D participants launched by Threat Models D and F  \\
  \(N_B\)              & number of Type-B participants launched by Threat Models D and F \\
  \bottomrule
\end{tabular}
\end{center}
\end{minipage}
\end{table}
\subsubsection{Adverse Effect and Attack Cost of Threat Model A}
Based on Formula (\ref{equation-17}), for independently mischievous participants, if they attack a target participant \(p_i\) successfully, the cost must meet the following condition.
\begin{equation} \label{equation-18}
\sum\nolimits_{tr(p_u ,p_i ) \in R_{nH} } {m_{p_u p_i } T_G (p_u )} /\sum\nolimits_{tr(p_w ,p_i ) \in R_H } {m_{p_w p_i } T_G (p_w )}>1.
\end{equation}

We utilize raw direct trust to replace normalized direct trust \(m_{p_u p_i}\), and beta function-based expectation to interpret direct trust. Thus we can transform Formula (\ref{equation-18}) into:
\begin{equation} \label{equation-19}
\sum\nolimits_{tr(p_u ,p_i ) \in R_{nH} } {\frac{1}{{|tr(p_u ,p_i )| + 2}} \cdot T_G (p_u )}  > \sum\nolimits_{tr(p_w ,p_i ) \in R_H } {\frac{{|tr(p_w ,p_i )| + 1}}{{|tr(p_w ,p_i )| + 2}} \cdot T_G (p_w )}.
\end{equation}

As transaction increases, the beta function-based direct trust over a pair of good participants \(p_w\) and \(p_i\) will enlarge, we can see that via Formula (\ref{equation-19}). Inversely, the direct trust over a pair of good and mischievous participants \(p_u\) and \(p_i\) will decline. Thereby, the appropriate time to launch an attack is at the beginning period, otherwise the mischievous participants would pay more cost. Thus, we calculate the adverse effect and attack cost when the transaction is performed only one-time, and suppose the feedback employs binary rating. In addition, we assume the number of independently mischievous participants is \(N_{nH}\) with the network level average trust \(T_{net}^M\), the number of good participants is \(N_H\) with the network level average trust \(T_{net}^G\). Thus, we can rewrite the condition (\ref{equation-19}) as:
\begin{equation}
N_{nH}  \cdot \frac{1}{{1 + 2}} \cdot T_{net}^M  > N_H  \cdot \frac{{1 + 1}}{{1 + 2}} \cdot T_{net}^G.
\end{equation}

Accordingly, the number of mischievous participants is \(N_{nH}\)=\(\left( {\left\lfloor {2N_H  \times T_{net}^G /T_{net}^M } \right\rfloor  + 1} \right)\). Given the independency of mischievous participants, each needs to launch at least one-time non-creditable rating to the target participant, thus the total non-creditable ratings are at least \(N_{nR}\)=\(\left( {\left\lfloor {2N_H  \times T_{net}^G /T_{net}^M } \right\rfloor  + 1} \right)\).
\subsubsection{Adverse Effect and Attack Cost of Threat Model B}
The mischievous participants organize a chain, and each mischievous participant in the chain would offer a high dishonest rating (1.0) to its partner. Nevertheless, since these mischievous participants cannot provide authentic services, they hardly gain positive ratings from good participants, this implies the chain-reinforced function in fact loses the trust transitivity effect. Thus, we have the following attack success condition.
\begin{equation}
\sum\nolimits_{tr(p_u ,p_i ) \in R_{dH} } {m_{p_u p_i } T_G (p_u )} /\sum\nolimits_{tr(p_w ,p_i ) \in R_H } {m_{p_w p_i } T_G (p_w )}>1.
\end{equation}

Accordingly, we have that:
\begin{equation}
\label{equation-22}
N_{dH}  \cdot \frac{1}{{1 + 2}} \cdot T_{net}^M  > N_H  \cdot \frac{{1 + 1}}{{1 + 2}} \cdot T_{net}^G.
\end{equation}

Thereby, the number of mischievous participants is \(N_{dH}\)=\(\left( {\left\lfloor {2N_H  \times T_{net}^G /T_{net}^M } \right\rfloor +1} \right)\). Besides dishonest (negative) ratings to target participant, the mischievous participants need give their chain-based partners dishonest (positive) ratings, the total dishonest ratings are \(N_{dR}\)=\(2 \cdot \left( {\left\lfloor {2N_H \times T_{net}^G /T_{net}^M } \right\rfloor +1} \right)\).
\subsubsection{Adverse Effect and Attack Cost of Threat Model C}
The camouflage participants not only form the reinforced trust-transitivity chain, but they can also provide authentic services with probability \(f\) to gain positive ratings from good participants. We assume the amount of authentic services provided by one camouflage participant is \(I_H\) and simultaneously it receives \(I_H\) positive ratings from good participants. Thus, we aggregate trust ingredient through received positive ratings:
\begin{equation}
\label{equation-23}
\begin{array}{l}
 T_{ti} (p_c ) = \sum\nolimits_{tr(p_w ,p_c ) \in R_H } {\frac{{|tr(p_w ,p_c )| + 1}}{{|tr(p_w ,p_c )| + 2}} \cdot T_G (p_w )}  \\
 \begin{array}{*{20}c}
   {} & {\begin{array}{*{20}c}
   {} &  =   \\
\end{array}}  \\
\end{array}I_H  \cdot \frac{{1 + 1}}{{1 + 2}} \cdot T_{net}^G  \\
 \end{array},
\end{equation}
where \(T_{ti}(p_c)\) denotes the trust ingredient of camouflage participant \(p_c\). We assume the number of camouflage participants is \(N_C\), they in return use gained trust ingredient \(T_{ti} (p_c )\) as distrust ingredient to attack target participant. Thus we can replace condition (\ref{equation-22}) as:
\begin{equation}
\label{equation-24}
\begin{array}{l}
 N_C  \cdot \frac{1}{{1 + 2}} \cdot I_H  \cdot \frac{{1 + 1}}{{1 + 2}} \cdot T_{net}^G  > N_H  \cdot \frac{{1 + 1}}{{1 + 2}} \cdot T_{net}^G  \\
 \begin{array}{*{20}c}
   {\begin{array}{*{20}c}
   {} & {}  \\
\end{array}} & {}  \\
\end{array}N_C  > 3 \cdot N_H /I_H  \\
 \end{array}.
\end{equation}

Given the direct trust over the chain is high (1.0), which implies the camouflage participants as a community only need one member to provide authentic services to gain positive ratings, this member in return can propagate its gained trust to its partner along the chain, by analogy, all camouflage participants can get the same trust value through the chain-based direct trust. Therefore, the attack cost includes: i) the number of camouflage participants is \(\left( {\left\lfloor {3N_H /I_H } \right\rfloor  + 1} \right)\); ii) the amount of authentic services provided by camouflage participants is \(I_H\); iii) the dishonest ratings to target participant are \(\left( {\left\lfloor {3N_H /I_H } \right\rfloor  + 1} \right)\); and iv) as well as the dishonest ratings to partners over the chain are \(\left( {\left\lfloor {3N_H /I_H } \right\rfloor + 1} \right)\).
\subsubsection{Adverse Effect and Attack Cost of Threat Model D}
We assume each Type-D participant provides \(I_H\) authentic services and receives \(I_H\) positive ratings from good participants. According to Formula (\ref{equation-23}), we have each Type-D participant's trust ingredient:
\begin{equation}
\begin{array}{l}
 T_{ti} (p_D ) = \sum\nolimits_{tr(p_w ,p_c ) \in R_H } {\frac{{|tr(p_w ,p_d )| + 1}}{{|tr(p_w ,p_d )| + 2}} \cdot T_G (p_w )}  \\
 \begin{array}{*{20}c}
   {} & {\begin{array}{*{20}c}
   {} &  =   \\
\end{array}}  \\
\end{array}I_H  \cdot \frac{{1 + 1}}{{1 + 2}} \cdot T_{net}^G  \\
 \end{array},
\end{equation}
where \(T_{ti}(p_D)\) denotes the trust ingredient of Type-D participant aggregated through gained positive ratings. Given Type-D participants cooperate with Type-B participants, i.e. Type-D participants proportionally give each Type-B participant direct trust 1/\(|N_B|\). Thus, each Type-B participant's trust ingredient can be calculated as:
\begin{equation}
\begin{array}{l}
 T_{ti} (p_B ) = N_D  \cdot \frac{1}{{N_B }} \cdot T_{ti} (P_D ) \\
 \begin{array}{*{20}c}
   {} & {\begin{array}{*{20}c}
   {} &  =   \\
\end{array}}  \\
\end{array}N_D  \cdot \frac{1}{{N_B }} \cdot I_H  \cdot \frac{{1 + 1}}{{1 + 2}} \cdot T_{net}^G  \\
 \end{array},
\end{equation}
where \(N_D\) and \(N_B\) (\(\ne\)0) stand for the numbers of Type-D and Type-B participants. To achieve attacking target, it must meet the following condition:
\begin{equation}
\label{equation-27}
\begin{array}{l}
 \begin{array}{*{20}c}
   {} & {} & {}  \\
\end{array}\begin{array}{*{20}c}
   {} & {}  \\
\end{array}\frac{{N_D }}{{1 + 2}} \cdot T_{ti} (p_D ) + \frac{{N_B }}{{1 + 2}} \cdot T_{ti} (p_B ) > \frac{{2N_H }}{{1 + 2}} \cdot T_{net}^G  \\
 N_D  \cdot \frac{1}{{1 + 2}} \cdot I_H  \cdot \frac{{1 + 1}}{{1 + 2}} \cdot T_{net}^G  + N_B  \cdot \frac{1}{{1 + 2}} \cdot N_D  \cdot \frac{1}{{N_B }} \cdot I_H  \cdot \frac{{1 + 1}}{{1 + 2}} \cdot T_{net}^G  > N_H  \cdot \frac{{1 + 1}}{{1 + 2}} \cdot T_{net}^G  \\
 \begin{array}{*{20}c}
   {} & {} & {} & {}  \\
\end{array}\begin{array}{*{20}c}
   {\begin{array}{*{20}c}
   {} & {}  \\
\end{array}} & {}  \\
\end{array}\begin{array}{*{20}c}
   {} & {}  \\
\end{array}N_D  > \frac{{3N_H }}{{2I_H }} \\
 \end{array}.
\end{equation}

Therefore, the attack cost includes: i) Type-D participants' amount is \(\left( {\left\lfloor {3N_H /2I_H } \right\rfloor  + 1} \right)\), Type-B participants' amount is \(N_B\); ii) Type-D participants need to altogether provide good participants \(\left( {\left( {\left\lfloor {3N_H /2I_H } \right\rfloor  + 1} \right) \cdot I_H } \right)\) authentic services; iii) offer \(\left( {\left\lfloor {3N_H /2I_H } \right\rfloor  + 1} \right) \cdot N_B\) dishonest ratings simultaneously to Type-B participants; and iv) Type-D and Type-B individuals need offer \(\left( {\left( {\left\lfloor {3N_H /2I_H } \right\rfloor  + 1} \right) + N_B } \right)\) dishonest ratings to target participant. From Formula (\ref{equation-27}), we can see that the number of Type-B participants does not affect the adverse effect of Type-D individuals, they just can be viewed as a trust transitivity bridge to receive Type-D participants' trust ingredient to further perform dishonest transactions.
\subsubsection{Adverse Effect and Attack Cost of Threat Model E}
The camouflage participants not only provide authentic services with probability \(f\) to gain positive ratings as feedback receivers, but also offer honest ratings with probability \(\eta\) as feedback raters. According to Formula (\ref{equation-23}), we know the camouflage participants can gain the trust ingredient \(T_{ti} (p_c)\) by contributing \(I_H\) honest transactions. Since they offer honest ratings with probability \(\eta\), which means they will propagate trust ingredient to good participants with proportion \(\eta\), and trust ingredient to the chain-based mischievous partners with probability (1-\(\eta\)). For the sake of explicitly understanding, we identify the camouflage participants in the chain as {\(p_{c_1}\), \(p_{c_2}\), \(\cdots\), \(p_{c_{Nc}}\)}. Might as well assume the 1st camouflage participant has provided \(I_H\) authentic services and gained trust ingredient \(T_{ti} (p_{c_1 } )\) already. Thus, with the trust transitivity the \(i\)th camouflage participant's trust ingredient can be calculated as:
\begin{equation}
T_{ti} (p_{c_i } ) = T_{ti} (p_{c_1 } ) \cdot (1 - \eta )^{(i - 1)}.
\end{equation}

To meet the attack success condition, we can rewrite the Formula (\ref{equation-24}) as:
\begin{equation}
\begin{array}{l}
  \quad \frac{1}{{1 + 2}} \cdot T_{ti} (p_{c_1 } ) + \frac{1}{{1 + 2}} \cdot T_{ti} (p_{c_2 } ) +  \cdots  + \frac{1}{{1 + 2}} \cdot T_{ti} (p_{c_{N_C } } ) > N_H  \cdot \frac{{1 + 1}}{{1 + 2}} \cdot T_{net}^G  \\
 \begin{array}{*{20}c}
   {} & {\frac{1}{{1 + 2}} \cdot I_H  \cdot \frac{{1 + 1}}{{1 + 2}} \cdot T_{net}^G  \cdot (1 + (1 - \eta ) +  \cdots  + (1 - \eta )^{(N_C  - 1)} ) > N_H  \cdot \frac{{1 + 1}}{{1 + 2}} \cdot T_{net}^G }  \\
\end{array} \\
 \begin{array}{*{20}c}
   {} & {} & {} & {}  \\
\end{array}\begin{array}{*{20}c}
   {} & {}  \\
\end{array}\begin{array}{*{20}c}
   {} & {N_C  > \log _{(1 - \eta )} (1 - \frac{{3N_H  \cdot \eta }}{{I_H }})}  \\
\end{array} \\
 \end{array}.
\end{equation}

Alike Threat Model C, only one camouflage participant needs to contribute authentic services, the others can gain trust ingredient through trust transitivity. The attack cost includes: i) the number of camouflage participants is \(\left( {\left\lfloor {\log _{(1 - \eta )} (1 - 3N_H  \cdot \eta /I_H )} \right\rfloor  + 1} \right)\); ii) the number of authentic services is \(I_H\); iii) the amount of dishonest ratings given to target participant is \(\left( {\left\lfloor {\log _{(1 - \eta )} (1 - 3N_H  \cdot \eta /I_H )} \right\rfloor  + 1} \right)\); iv) the dishonest ratings given to mischievous partners are \(\left( {\left\lfloor {\log _{(1 - \eta )} (1 - 3N_H  \cdot \eta /I_H )} \right\rfloor  + 1} \right)\);  and v) given the amount of dishonest ratings to the mischievous partners and attack target, in addition to the probability \(\eta\), we can derive the total ratings offered by camouflage participants are \(\left( {\left\lfloor {\frac{{2\left( {\left\lfloor {\log _{(1 - \eta )} (1 - 3N_H  \cdot \eta /I_H )} \right\rfloor  + 1} \right)}}{{(1 - \eta )}}} \right\rfloor  + 1} \right)\), the number of honest ratings is \(\left( {\left\lfloor {\frac{{2\eta  \cdot \left( {\left\lfloor {\log _{(1 - \eta )} (1 - 3N_H  \cdot \eta /I_H )} \right\rfloor  + 1} \right)}}{{(1 - \eta )}}} \right\rfloor  + 1} \right)\).
\subsubsection{Adverse Effect and Attack Cost of Threat Model F}
The honest ratings offered by spy participants mainly influence the trust transitivity from Type-D to Type-B participants, i.e. the Type-D participants aim to split the gained trust ingredient from good participants, in return to rate back to the good ones honestly. In this way, the trust ingredient propagated to Type-B participants from Type-D participants declines by probability \(\gamma\). Taking into account this, we redefine the trust gradient for each Type-B participant as:
\begin{equation}
\begin{array}{l}
 T_{ti} (p_B ) = N_D  \cdot \frac{1}{{N_B }} \cdot (1 - \gamma ) \cdot T_{ti} (P_D ) \\
 \begin{array}{*{20}c}
   {} & {\begin{array}{*{20}c}
   {} &  =   \\
\end{array}}  \\
\end{array}N_D  \cdot \frac{1}{{N_B }} \cdot (1 - \gamma ) \cdot I_H  \cdot \frac{{1 + 1}}{{1 + 2}} \cdot T_{net}^G  \\
 \end{array}.
\end{equation}

In accordance, we rewrite the attack success condition (\ref{equation-27}) as:
\begin{equation}
\begin{array}{l}
 \begin{array}{*{20}c}
   {} & {} & {}  \\
\end{array}N_D  \cdot \frac{1}{{1 + 2}} \cdot T_{ti} (p_D ) + N_B  \cdot \frac{1}{{1 + 2}} \cdot T_{ti} (p_B ) > N_H  \cdot \frac{{1 + 1}}{{1 + 2}} \cdot T_{net}^G  \\
 N_D  \cdot \frac{1}{{1 + 2}} \cdot I_H  \cdot \frac{{1 + 1}}{{1 + 2}} \cdot T_{net}^G  + N_B  \cdot \frac{1}{{1 + 2}} \cdot N_D  \cdot \frac{1}{{N_B }} \cdot (1 - \gamma ) \cdot I_H  \cdot \frac{{1 + 1}}{{1 + 2}} \cdot T_{net}^G  > N_H  \cdot \frac{{1 + 1}}{{1 + 2}} \cdot T_{net}^G  \\
 \begin{array}{*{20}c}
   {} & {} & {} & {}  \\
\end{array}\begin{array}{*{20}c}
   {\begin{array}{*{20}c}
   {} & {}  \\
\end{array}} & {}  \\
\end{array}\begin{array}{*{20}c}
   {} & {}  \\
\end{array}N_D  > \frac{{3N_H }}{{(2 - \gamma ) \cdot I_H }} \\
 \end{array}.
\end{equation}

Therefore, the attack cost includes: i) Type-D participants' amount is \(\left( {\left\lfloor {\frac{{3N_H }}{{(2 - \gamma ) \cdot I_H }}} \right\rfloor  + 1} \right)\), Type-B participants' amount is \(N_B\); ii) Type-D participants need offer \(\left( {\left\lfloor {\frac{{3N_H }}{{(2 - \gamma ) \cdot I_H }}} \right\rfloor  + 1} \right) \cdot I_H\) authentic services; and iii) give \(\left( {\left\lfloor {\frac{{3N_H }}{{(2 - \gamma ) \cdot I_H }}} \right\rfloor  + 1} \right) \cdot N_B\) dishonest ratings to Type-B participants; iv) Type-D and Type-B participants need offer \(\left( {\left( {\left\lfloor {\frac{{3N_H }}{{(2 - \gamma ) \cdot I_H }}} \right\rfloor  + 1} \right) + N_B } \right)\) dishonest ratings to target participant; v) based on the dishonest ratings given to attack target and Type-B participants, referring to probability \(\gamma\) we can infer the total ratings offered by Type-D participants are \(\left( {\left\lfloor {\left( {\left\lfloor {\frac{{3N_H }}{{(2 - \gamma ) \cdot I_H }}} \right\rfloor  + 1} \right) \cdot (1 + N_B )/(1 - \gamma )} \right\rfloor  + 1} \right)\), the honest ratings are \(\left( {\left\lfloor {\left( {\left\lfloor {\frac{{3N_H }}{{(2 - \gamma ) \cdot I_H }}} \right\rfloor  + 1} \right) \cdot (1 + N_B ) \cdot \gamma /(1 - \gamma )} \right\rfloor  + 1} \right)\). Alike Threat Model D, the number of Type-B participants does not affect the adverse effect of Type-D participants, they just receive Type-D participants' trust ingredient for launching more mischievous transactions.
\subsection{Attack Behavior Analysis and Evaluation}
We run a group of experiments to further analyze the six attack behaviors referring to Table \ref{table:two-0}, wherein "TM" denotes threat models. The attack behavior is deeply analyzed through observing the number (\#) of mischievous participants and the number (\#) of honest/dishonest ratings with varying SMPs' authentic services.
\begin{table}
\caption{Variables and parameters}
\label{table:two-0}
\begin{minipage}{\columnwidth}
\begin{centering}
\renewcommand{\multirowsetup}{\centering}
\begin{tabular}{lclclclclclclclclclclclclcl}
  \toprule
  \multirow{1}{0.35cm}{TM} & \multirow{1}{0.38cm}{\(N_{nH}\)} & \multirow{1}{0.38cm}{\(N_{dH}\)} & \multirow{1}{0.8cm}{\(N_H\)} &  \multirow{1}{0.38cm}{\(R_{nH}\)} &  \multirow{1}{0.38cm}{\(R_{dH}\)}  &  \multirow{1}{0.35cm}{\(T_{net}^M\)} & \multirow{1}{1.0cm}{\(T_{net}^G\)} & \multirow{1}{0.35cm}{\(N_C\)} & \multirow{1}{0.65cm}{\(I_H\)} & \multirow{1}{0.35cm}{\(N_B\)} & \multirow{1}{0.35cm}{\(N_D\)} & \multirow{1}{0.35cm}{\(\eta\)} & \multirow{1}{0.35cm}{\(\gamma\)} \\
\hline
\multicolumn{1}{c}{A} & v & / & h(1-19)   & v & /  & 0.35 & 0.75, 0.85, 0.95 & / & /       & / & / & /   & / \\
\multicolumn{1}{c}{B} & / & v & h(1-19)   & / & v  & 0.35 & 0.75, 0.85, 0.95 & / & /       & / & / & /   & / \\
\multicolumn{1}{c}{C} & / & / & 5, 10, 15 & / & v  & /    & /                & v & h(1-19) & / & / & /   & / \\
\multicolumn{1}{c}{D} & / & / & 5, 10, 15 & / & v  & /    & /                & / & h(1-19) & v & v & /   & / \\
\multicolumn{1}{c}{E} & / & / & 5, 10, 15 & / & v  & /    & /                & v & h(1-19) & / & / & 0.2 & / \\
\multicolumn{1}{c}{F} & / & / & 5, 10, 15 & / & v  & /    & /                & v & h(1-19) & v & v & /   & 0.2 \\
\hline
\multicolumn{14}{l}{v--observed value (vertical axis), h--observed value (horizon axis), /--variable inapplicable.}  \\
\bottomrule
\end{tabular}
\end{centering}
\end{minipage}
\end{table}

\begin{figure}[bt]
  \includegraphics [width=4.2in,height=6.7in]{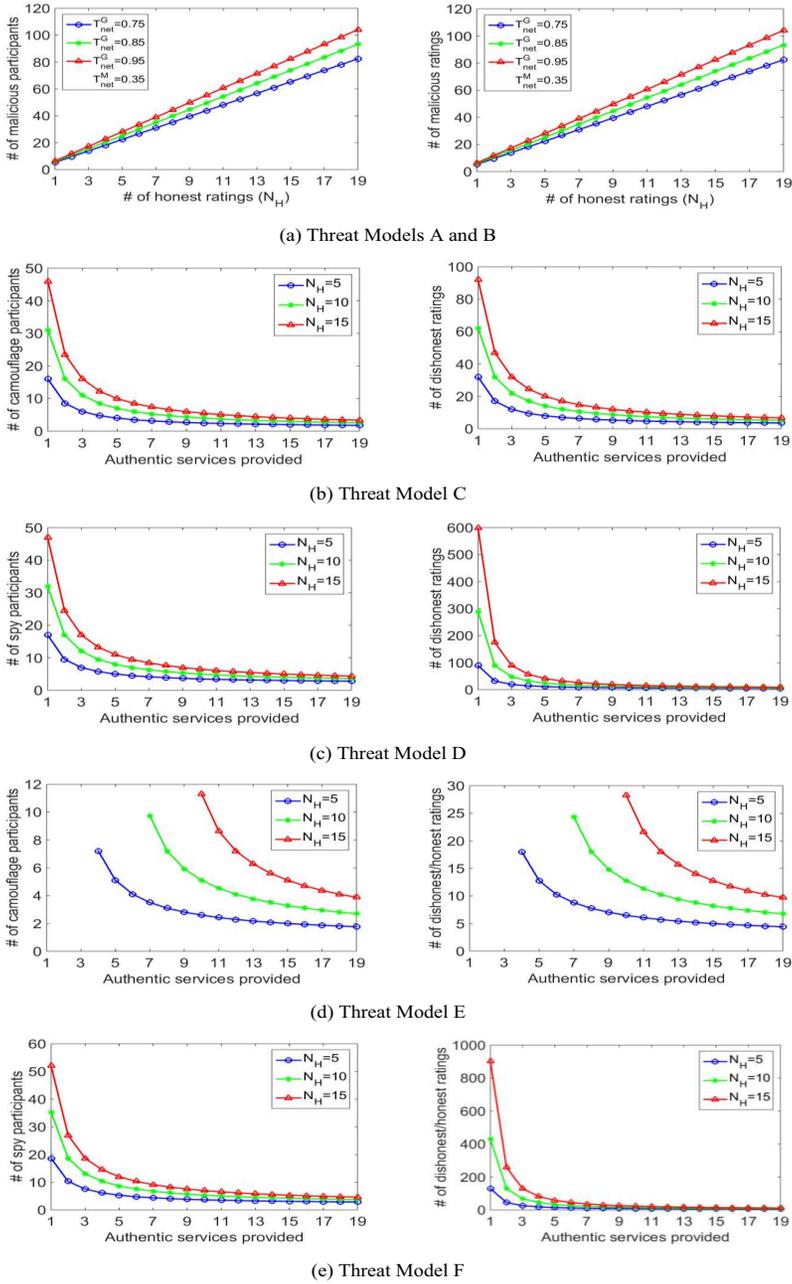}
  \caption{Attack cost evaluation.}
  \label{fig1}
\end{figure}

Fig. \ref{fig1} shows the attack costs under Threat Models A-F. We can observe several interesting and reasonable phenomena: i) for Threat Models A and B, since the mischievous participants cannot offer authentic services to gain trust ingredient, the \# of mischievous participants needed goes up linearly as the honest ratings given by good participants increase, as well as the \# of dishonest ratings enlarges linearly; ii) as the \# of authentic services enlarges in Threat Model C, both camouflage participants and dishonest ratings decline gradually. Due to the reinforced trust transitivity of chain-based camouflage participants, it only needs one camouflage participant to provide authentic services. The camouflage participants not only need provide dishonest ratings to target participant, but they also need provide dishonest ratings to their partners along the chain, hence the dishonest ratings are twice that of camouflage participants; iii) for Threat Model D, we set Type-D participants and Type-B participants equally, although the amount of Type-B participants does not affect the "trust ingredient" of Type-D group. We can observe from Fig. \ref{fig1}(c) that the spy participants and dishonest ratings have a declining tendency as authentic services uploaded by Type-D participants increase. However, apart from both Type-D and Type-B individuals give target participant dishonest ratings, each Type-D participant would give all Type-B participants dishonest (positive) ratings to promote trust ingredient. Therefore, the dishonest ratings are much more than that in Threat Model C in which each camouflage participant only needs one dishonest rating to build the chain. In addition, all Type-D participants need provide \(I_H\) authentic services rather than only needing one individual. Thereby, the attack cost in Threat Model D is much more than that in Threat Model C; iv) since the exponential function based trust propagation deriving from the honest ratings from camouflage participants to good ones in Threat Model E in fact diminishes trust ingredient of the mischievous partners in the chain, this naturally requires more camouflage participants and dishonest/honest ratings compared with Threat Model C. Owing to \(\log _{(1 - \eta )} (1 - \frac{{3N_H  \cdot \eta }}{{I_H }})\) must be subject to 0\(<\)\(1 - \frac{{3N_H  \cdot \eta }}{{I_H }}\)\(<\)1 on the condition 0\(<\)(1-\(\eta\))\(<\)1, we have \(I_H\)\(>\)\(3N_H \cdot \eta\). Normally, since mischievous participants would not like to give honest ratings with high probability \(\eta\), we set \(\eta\)=0.2; v) since Threat Model F also exists "trust leakage", it needs more attack cost compared with Threat Model D. We set \(\gamma\)=0.2 as well.

Upon the observation and analysis above, we can conclude the SMPs can indeed decrease attack cost through providing authentic services. This also reveals the ground-truth that the mischievous participants would pay more cost if they occasionally behave honestly rather than purely badly to try to keep them undetected. This reflects the realistic behind reason why mischievous participants are not willing to provide authentic services.

The above deduces the attack cost in theory, we next verify our argument through performing a set of experiments using popular trust metrics: BetaTrust \cite{35-Josang}, EigenTrust \cite{32-Kamvar}, ServiceTrust \cite{33-Su} and ServiceTrust\(^{++}\) \cite{43-Su}. To fairly evaluate the attack cost, we set same experiment environment as reported in EigenTrust \cite{32-Kamvar}, i.e., the experiment platform includes 60 good participants and 40 mischievous participants. The total transaction number is set to be 10 times the system size, i.e. 1000 transactions. For Threat Model C, the probability \(f\) is set as 0.4; besides \(f\)=0.4, the probability \(\eta\) in Threat Model E is set as 0.2. In addition, the 40 mischievous participants are equally divided into Type-D and Type-B participants in Threat Model D, the probability \(\gamma\) is set as 0.2 likewise in Threat Model F.
\begin{figure}[bt]
  \includegraphics [width=4.6in,height=3.2in] {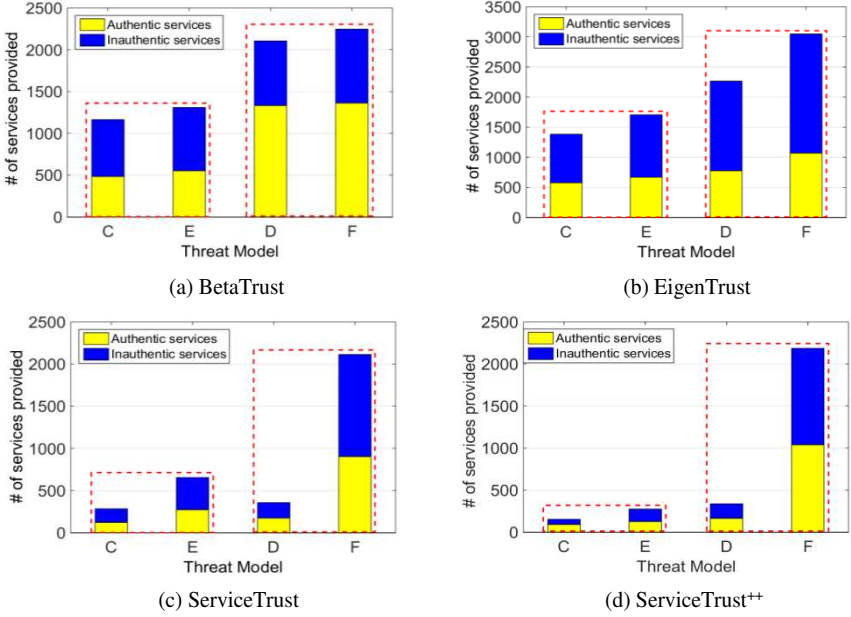} 
  \caption{Transactional behaviors of SMPs.}
  \label{fig2}
\end{figure}

From Fig. \ref{fig2}, we observe that: i) the camouflage and spy participants gain high trust ingredient through providing authentic services; on the other hand, they perform badly to offer inauthentic services through aggregated global trust; ii) compared with the SMPs in Threat Models C and D, the more-sophisticated mischievous participants in Threat Models E and F need to provide more authentic services. For example, the numbers of authentic services provided by Threat Models C and E are (483, 551) in BetaTrust, (578, 669) in EigenTrust, (122, 272) in ServiceTrust and (91, 125) in ServiceTrust\(^{++}\); the numbers of authentic services provided by Threat Models D and F are (1332, 1361), (777, 1070), (173, 902) and (164, 1039) respectively. This interprets the adverse effect of Threat Models E and F is naturally larger than that of Threat Models C and D, i.e. the more-sophisticated misbehavior participants own more opportunities to offer more inauthentic services. For instance, the numbers of inauthentic services provided by Threat Models C and E are (681, 758), (805, 1038), (162, 383) and (62, 149) in the four trust metrics; the numbers of inauthentic services provided by Threat Models D and F are (774, 886), (1488, 1978), (183, 1211) and (173, 1147). Therefore, as analyzed previously, Threat Models E and F indeed need more attack cost and simultaneously bring in more severe attack effect; iii) compared with the simple trust metrics BetaTrust \cite{35-Josang} and EigenTrust \cite{32-Kamvar}, the feedback credibility-based ServiceTrust \cite{33-Su} and threshold-controlled trust metric ServiceTrust\(^{++}\) \cite{43-Su} could have a much better performance, especially under Threat Models C and D, although they still suffer from the more-sophisticated misbehaviors in Threat Models E and F. Table \ref{table:two} summarizes and compares the six threat models in terms of attack cost, adverse effect, trust transitivity and defense strategy.
\begin{table}
\caption{Adverse behavior analysis and evaluation}
\label{table:two}
\begin{minipage}{\columnwidth}
\begin{centering}
\renewcommand{\multirowsetup}{\centering}
\begin{tabular}{lclclclclclcl}
  \toprule
  \multirow{1}{1cm}{Threat Model} & \multirow{1}{1.8cm}{Mischievous Participants} &  \multirow{1}{1.8cm}{Mischievous Ratings} & \multirow{1}{1.5cm}{Authentic Services} &  \multirow{1}{1cm}{Adverse Effect} &  \multirow{1}{1.8cm}{Trust Transitivity}  &  \multirow{1}{1cm}{Defense strategy} \\
      &         &            &        &           &            &     \\
\hline
  \multicolumn{1}{c}{A}   & linear  & \multicolumn{1}{c}{linear}	 & none	  & \multicolumn{1}{c}{weak}      & none       & \multicolumn{1}{c}{easy} \\
  \multicolumn{1}{c}{B}   & linear  & \multicolumn{1}{c}{linear}	 & none	  & \multicolumn{1}{c}{weak}      & none       & \multicolumn{1}{c}{easy} \\
  \multicolumn{1}{c}{C}   & medium	& \multicolumn{1}{c}{small} 	 & small  & \multicolumn{1}{c}{mediocre}  &	 existence  & \multicolumn{1}{c}{solvable}\\
  \multicolumn{1}{c}{D}   & medium	& \multicolumn{1}{c}{large}	     & large  & \multicolumn{1}{c}{mediocre}  & existence  & \multicolumn{1}{c}{solvable} \\
  \multicolumn{1}{c}{E}   &	medium 	& \multicolumn{1}{c}{small} 	 & small  & \multicolumn{1}{c}{strong} 	  & existence  & \multicolumn{1}{c}{difficult} \\
  \multicolumn{1}{c}{F}	  & medium	& \multicolumn{1}{c}{large} 	 & large  & \multicolumn{1}{c}{strong}	  & existence  & \multicolumn{1}{c}{difficult} \\
  \bottomrule
\end{tabular}
\end{centering}
\end{minipage}
\end{table}
\section{VULNERABILITY ANALYSIS OF TRUST AGGREGATION MODELS} \label{sec:five}
\subsection{Trust Aggregation Principle}
Trust aggregation in a decentralized network can be deemed as the fusion of feedback information over a graph organized by various nodes (participants). It is naturally subject to two factors: i) pairwise direct trust with local trust aggregation; and ii) trust propagation kernel-conducted global trust aggregation. Apart from UDTP and TCTP, we add non-propagation (NP) kernel, i.e. the global trust of a participant \(p_i\) is aggregated through the direct trust placed on \(p_i\) from the neighboring participants plus the recommended trust by other participants. Thereby, trust metrics can be routinely categorized as six combinations with respect to two direct trust aggregation fashions and three trust propagation kernels.

\begin{definition} [Raw Direct Trust with Non-Propagation Kernel, RNP] Trust is aggregated using the raw direct trust inferred from pairwise positive and negative ratings, the trust in fact only captures 1-hop feedback information without trust propagation.
\end{definition}

\begin{definition} [Feedback Credibility-Weighted Direct Trust with Non-Propagation Kernel, CNP] Trust is aggregated using FCW direct trust referring to feedback rater level credibility or feedback rating score level credibility, the trust only captures 1-hop feedback formation without trust propagation.
\end{definition}

\begin{definition} [Raw Direct Trust with Uniformly Distributed Trust Propagation Kernel, RUDP] Trust is aggregated using the raw direct trust inferred from pairwise positive and negative ratings, the trust captures \(k\)-hop (network-horizon) feedback information through UDTP kernel.
\end{definition}

\begin{definition} [Feedback Credibility-Weighted Direct Trust with Uniformly Distributed Trust Propagation Kernel, CUDP] Trust is aggregated using FCW direct trust referring to feedback rater level credibility or feedback rating score level credibility, the trust captures \(k\)-hop (network-horizon) feedback information through UDTP kernel.
\end{definition}

\begin{definition} [Raw Direct Trust with Threshold-Controlled Trust Propagation Kernel, RTCP] Trust is aggregated using the raw direct trust inferred from pairwise positive and negative ratings, the trust in fact captures partial intended feedback information through TCTP kernel.
\end{definition}

\begin{definition} [Feedback Credibility-Weighted Direct trust with Threshold-Controlled Trust Propagation Kernel, CTCP] Trust is aggregated using FCW direct trust referring to feedback rater level credibility or feedback rating score level credibility, the trust in fact captures partial intended feedback information through TCTP kernel.
\end{definition}
\subsection{Attack Analysis of Reference Aggregation Models}
\begin{figure}[bt]
  \includegraphics [width=4.4in,height=1.4in] {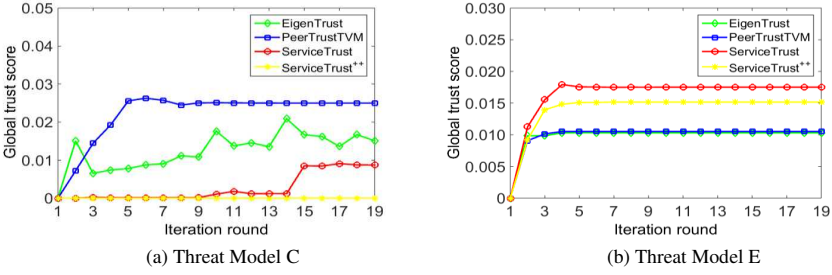} 
  \caption{Global trust of camouflage participants.}
  \label{fig3}
\end{figure}

To deeply study the pros and cons of reference trust aggregation models, we primarily select four representative trust metrics to launch a set of experiments with two strategic Threat Models C and E to exhibit how trust changes as iteration round increases: i) RUDP trust metric-EigenTrust \cite{32-Kamvar}; ii) feedback rater level credibility-based CUDP trust metric-PeerTrustTVM \cite{38-Xiong}; iii) feedback rating score level credibility-based CUDP trust metric-ServiceTrust \cite{33-Su}; iv) feedback rating score level credibility-based CTCP trust metric-ServiceTrust\(^{++}\) \cite{43-Su}. In our experiments, we set the numbers of good, camouflage and pre-trusted participants as 60, 40 and 3, \(f\) as 0.4 and \(\eta\) as 0.5. Fig. \ref{fig3} exhibits the global trust of camouflage participants.

We randomly choose one camouflage participant to observe the variation tendency of global trust with increasing the number of iteration rounds. In the presence of the Threat Model C, we observe that: i) in EigenTrust \cite{32-Kamvar}, the camouflage participant's global trust enlarges gradually as the number of iterations increases; ii) in PeerTrustTVM \cite{38-Xiong}, the global trust becomes large shortly within 5 iteration rounds. In contrast, in ServiceTrust \cite{33-Su}, the global trust goes up slowly. This is because PeerTrustTVM \cite{38-Xiong} can promote the trust propagation from a camouflage participant to its partner along the chain, deriving from the high self-trust based feedback credibility weighted raw direct trust. In comparison, ServiceTrust \cite{33-Su} dramatically reduces the trust propagation from good participants to camouflage ones by leveraging the dissimilarity between good and camouflage participants. This indicates that the rating similarity based credibility combined with the dissimilarity based trust decaying in ServiceTrust \cite{33-Su} is an effective mechanism to constrain the dishonest ratings to propagate in the presence of camouflage participants; iii) in ServiceTrust\(^{++}\) \cite{43-Su}, the global trust is always zero regardless what the specific iteration round is, this implies that the TCTP kernel succeeds in cutting off trust transitivity paths/edges from good participants to camouflage ones. The primary improvement of ServiceTrust\(^{++}\)~\cite{43-Su} over ServiceTrust \cite{33-Su} is the trust propagation kernel. Put differently, the UDTP kernel based ServiceTrust \cite{33-Su} cannot prevent the camouflage participants obtaining positive global trust, but the TCTP kernel based ServiceTrust\(^{++}\) \cite{43-Su} can throughout block trust propagation through the system-inferred threshold.

From the results of Threat Model E, we observe the four representative trust metrics suffer from this more-sophisticated attack behavior, even the feedback credibility-based trust metrics and TCTP kernel-conducted trust metrics all become ineffective. The root-cause behind lies in Threat Model E invalidates both FCW direct trust and TCTP kernel through making the good in appearance but more-sophisticated mischievous participants imitate good participants as alike as possible. In essence, the radical reasons as analyzed in work \cite{4-Fan} lies in the pairwise similarity has become inadequate to differentiate camouflage participants from good participants.

Next, we exhibit the inherent vulnerabilities through analyzing the attack behaviors of reference aggregation models in conjunction with transactional performance. We launch six groups of experiments to study the fraction of failed transactions using several referral trust metrics: i) random trust metric-NoneTrust; ii) RNP trust metric-BetaTrust \cite{35-Josang}; iii) RTCP trust metric-AdaptiveTrust \cite{42-Chen}, here we can recognize it as a RTCP trust metric due to the setting of minimal honesty trust threshold (0.5) with time slot-based trust update; iv) RUDP trust metric-EigenTrust \cite{32-Kamvar}; v) CUDP trust metric-PeerTrustTVM \cite{38-Xiong}; vi) CNP trust metric-PeerTrustPSM \cite{38-Xiong}; vii) CUDP trust metric-ServiceTrust \cite{33-Su}; and viii) CTCP trust metric-ServiceTrust\(^{++}\) \cite{43-Su}.

\begin{figure}[bt]
  \includegraphics [width=5.2in,height=4.2in] {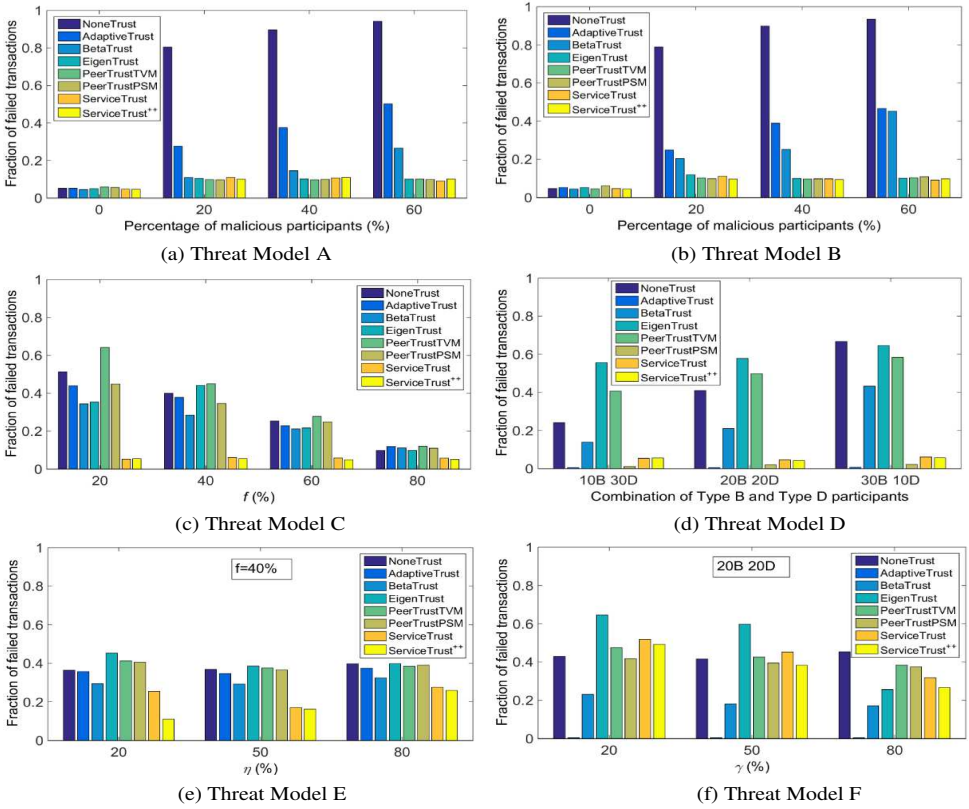}
  \caption{Performance with Threat Models A-F.}
  \label{fig4}
\end{figure}

To keep experiment configuration identical, we set same environment as EigenTrust \cite{32-Kamvar}, i.e., the experiment platform has 100 participants and 3 pre-trusted participants. In Threat Models A and B, the percentage of mischievous participants varies from 0 to 60\%. In Threat Model C, the \(f\) increases from 20\% to 80\%. In Threat Model D, 40 spy participants organize three combinations: (10 Type-B, 30 Type-D), (20 Type-B, 20 Type-D), (30 Type-B, 10 Type-D). In Threat Model E, we set the number of camouflage participants as 40 and keep \(f\)=0.4, then vary \(\eta\) from 0.2 to 0.8. In Threat Model F, we divide the spy participants into 20 Type-B and 20 Type-D, then change \(\gamma\) from 0.2 to 0.8. Fig. \ref{fig4} exhibits the performance.

For Threat Models A and B, most trust metrics are effective deriving from the zero-value direct trust from good individuals to mischievous ones, which leads to zero global trust for mischievous participants, such as EigenTrust \cite{32-Kamvar}, PeerTrustTVM \cite{38-Xiong}, PeerTrustPSM \cite{38-Xiong}, ServiceTrust \cite{33-Su} and ServiceTrust\(^{++}\) \cite{43-Su}. However, apart from the random trust metric NoneTrust which randomly selects transacted participants, the RTCP trust metric AdaptiveTrust \cite{42-Chen} and RNP trust metric BetaTrust \cite{35-Josang} cannot conquer the two simple attacks. In BetaTrust \cite{35-Josang}, since it derives trust using the difference between successful and unsuccessful transactions without normalizing the direct trust into the interval [0, 1], thus, the mischievous participants can obtain non-zero trust scores. Consequently, some mischievous participants might be selected as transacted targets according to the probabilistic selection criterion. For AdaptiveTrust \cite{42-Chen}, it assumes each participant's initial trust score as 0.5, thus the mischievous participants can also be selected as transacted targets with a big probability, especially in the beginning stage the difference of most participants' trust is subtle.

For Threat Model C, the random, RNP, CNP, CUDP and RUDP trust metrics all become invalid due to the existence of strategically mischievous behaviors of the camouflage participants, which are acting as good participants at a certain probability \(f\). For the CUDP trust metric PeerTrustTVM \cite{38-Xiong} and CNP trust metric PeerTrustPSM \cite{38-Xiong}, they behave poorly since both the feedback rater level credibility and feedback rating score level credibility cannot effectively prevent good participants to give direct trust on camouflage participants. However, ServiceTrust \cite{33-Su} uses the CUDP trust metric by employing the feedback rating score level credibility to weight direct trust, which effectively decreases the direct trust from good participants to camouflage participants. In addition, with the CTCP trust metric, ServiceTrust\(^{++}\) \cite{43-Su} can effectively cut off dishonest trust propagation to camouflage individuals even though the direct trust weighted by feedback rating score level credibility is not dropped to the ground-truth level (zero). Hence, the CTCP trust metric can effectively conquer strategic camouflage attack.

For Threat Model D, AdaptiveTrust \cite{42-Chen} conquers, this is because there exists an amending mechanism, i.e. once a participant finds another transacted participant gives a bad service, then it sets this participant's trust as 0.0. For CNP-based PeerTrustPSM \cite{38-Xiong} and CUDP-based ServiceTrust \cite{33-Su}, they can conquer this kind of spy attack since the similarity-inferred feedback credibility approaches are effective to decline the trust of purely mischievous Type-B participants deriving from the total dissimilarity between good participants and spy participants. In ServiveTrust\(^{++}\) \cite{43-Su}, both the feedback rating score level credibility and controlled trust propagation kernel doubly constrain these spy participants' trust aggregation. However, for the RNP-based BetaTrust \cite{35-Josang}, RUDP-based EigenTrust \cite{32-Kamvar} and CUDP-based PeerTrustTVM \cite{38-Xiong}, they all suffer from this spy attack. BetaTrust \cite{35-Josang} and EigenTrust \cite{32-Kamvar} would assign high trust scores to Type-D participants. Type-D participants can manipulate PeerTrustTVM \cite{38-Xiong} through the high feedback rater level credibility. In summary, the adequate feedback credibility, supported in PeerTrustPSM \cite{38-Xiong}, ServiceTrust \cite{33-Su} and ServiceTrust\(^{++}\) \cite{43-Su}, can effectively control and block the direct trust from good participants to spy participants.

For Threat Model E, all the trust metrics suffer from this more-sophisticated adversary behavior, even the CUDP-based ServiceTrust \cite{33-Su} and CTCP-based ServiceTrust\(^{++}\) \cite{43-Su} cannot resist in spite of a better performance compared with other trust metrics. This is because the tremendously similar transactional behavior between camouflage participants with probabilistically honest ratings and good participants makes the trust metrics become extremely difficult to distinguish these good in appearance but mischievous participants. Singly from the viewpoint of direct trust aggregation, it is hard to degrade the raw direct trust or FCW direct trust, not to mention the trust propagation kernels. Compared to the raw direct trust and inadequate FCW direct trust, such as AdaptiveTrust \cite{42-Chen}, BetaTrust \cite{35-Josang}, EigenTrust \cite{32-Kamvar} and PeerTrustTVM \cite{38-Xiong}, the adequate FCW direct trust can make the ratio of failed transactions decline to some extent, such as PeerTrustPSM \cite{38-Xiong}, ServiceTrust \cite{33-Su} and ServiceTrust\(^{++}\) \cite{43-Su}.

For Threat Model F, apart from AdaptiveTrust \cite{42-Chen} with amending mechanism, all other trust metrics suffer from this more-sophisticated spy misbehavior as well. The fundamental reason also lies in these spy participants behave almost alike as good participants to a large extend through adjusting the probability of offering honest ratings.
\subsection{Summary of Attack Analysis}
Upon the attack analysis above, we can conclude that: i) for independently and collectively malicious behaviors in Threat Models A and B, the raw direct trust or FCW direct trust can be an effective fashion as transaction enlarges, accordingly the aggregated trust levels can differentially represent good and mischievous participants using whatever NP, UDTP or TCTP kernels; ii) for strategic camouflage and spy behaviors in Threat Models C and D, the raw direct trust becomes ineffective. The adequate FCW direct trust becomes valid to a large extent but cannot throughout constrain the trust propagation from good to the camouflage and spy participants into ground-true level. If TCTP kernel can be employed to control how to propagate/block trust among different categories of participants, the trust metrics would be effective. That is to say a reliable trust metric needs to employ an adequate FCW direct trust plus an appropriate TCTP kernel to resist camouflage and spy attacks; iii) for more-sophisticated mischievous participants in Threat Models E and F, considering the extremely similar transactional behavior with good participants, it is hard to defend due to high similarity-inferred feedback credibility, in addition to the invalid similarity-inferred threshold. Thereby, even the adequate FCW direct trust and TCTP kernel cannot conquer these more-sophisticated attacks, but can decline the trust levels of camouflage and spy participants to an extent.

To further dig out the root-causes why diverse categories of trust metrics suffer from the more-sophisticated camouflage and spy attacks, we utilize the real-word interactive network-Epinions to launch a group of experiments to unveil the behind reasons. We add 30 nodes into the 100 regular/good nodes organized Epinions network, recognizing the 30 added nodes as strategically mischievous nodes with Threat Models E and F. For regular nodes, we straightly adopt Zipf distribution to assign pairwise edge weight (direct trust) for each pair of connected nodes. The regular nodes select a decimal from interval [0.85, 1.0] to rate the added camouflage nodes, and the added nodes select a decimal from interval [\(\eta\)-0.05, \(\eta\)+0.05] to rate regular nodes under Threat Model E.
\begin{figure}[bt]
  \includegraphics [width=5.2in,height=3.4in]{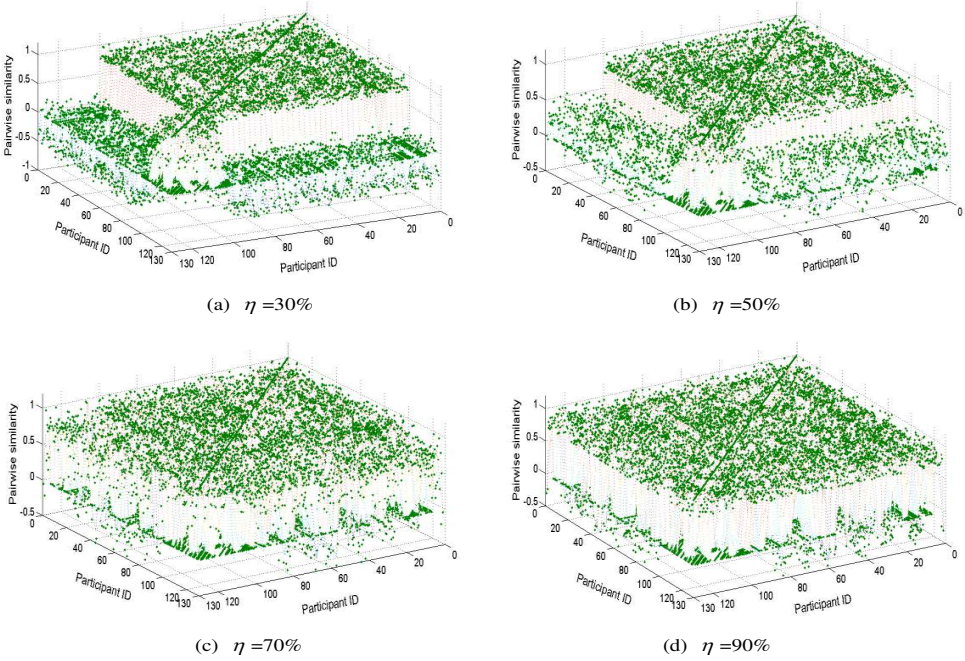}
  \caption{Pairwise Similarity under Threat Model E.}
  \label{fig5}
\end{figure}

Fig. \ref{fig5} exhibits the pairwise similarity among the 130 nodes wherein 100 nodes (ID=1-100) belong to regular nodes, 30 nodes (ID=101-130) pertain to mischievous nodes. The experimental results interpret the pairwise similarity between camouflage and good nodes gradually increases as the \(\eta\) enlarges from 30\% to 90\%. The similarity between camouflage and good nodes is almost in the same level between good nodes themselves when \(\eta\) is up to 90\%. This indicates when \(\eta\) is large enough the similarity-based feedback credibility becomes invalid to degrade raw direct trust from good to camouflage nodes, as well as the similarity-based threshold becomes invalid to block dishonest trust propagation owning to high similarity-based direct trust. Similar results can be also obtained under Threat Model F while extending the \(\gamma\) step by step. This reveals our proposed Threat Models E and F can indeed make the more-sophisticated mischievous participants behave extremely similarly as good ones, making them become extremely difficult to defend.

To explicitly illuminate the features of the state of the art trust metrics, we categorize them into six communities to sketch the weakness and attack resilience with respect to the six threat models in Tables \ref{table:three}-\ref{table:eight}. THR\_A-F denotes Threat Models A-F, "+" denotes trust metric can conquer threat model, "-" oppositely indicates trust metric cannot.
\begin{table}
\centering
\caption{Attack Analysis on RNP-based Trust Metrics}
\label{table:three}
\begin{minipage}{\columnwidth}
\renewcommand{\multirowsetup}{\centering}
\begin{tabular}{lclclclclclclcl}
  \toprule
  \multicolumn{1}{c|}{\multirow{2}{*}{Trust Metrics}} & \multicolumn{6}{|c}{Defense Countermeasure} \\
  \cline{2-7}
                   & \multicolumn{1}{|c|}{THR\_A}     & \multicolumn{1}{|c|}{THR\_B}  & \multicolumn{1}{|c|}{THR\_C}   & \multicolumn{1}{|c|}{THR\_D}     & \multicolumn{1}{|c|}{THR\_E}  & \multicolumn{1}{|c}{THR\_F}     \\
  \hline
  Marti et al. \citeyearpar{15-Marti}            & +  & \multicolumn{1}{c}{+}  & -  & \multicolumn{1}{c}{-}  & -  & \multicolumn{1}{c}{-}  \\
  J{\o}sang et al. \citeyearpar{55-Josang}          & +  & \multicolumn{1}{c}{+}  & -	 & \multicolumn{1}{c}{-}  & -  & \multicolumn{1}{c}{-}  \\
  TrustWalker \citeyearpar{21-Jamali}     &	+  & \multicolumn{1}{c}{+}  & +	 & \multicolumn{1}{c}{+}  & -  & \multicolumn{1}{c}{-}  \\
  Wang et al. \citeyearpar{30-Wang}                &	+  & \multicolumn{1}{c}{+}	& -	 & \multicolumn{1}{c}{-}  & -  & \multicolumn{1}{c}{-}  \\
  Liu et al. \citeyearpar{23-Liu}	              & +  & \multicolumn{1}{c}{+}	& +	 & \multicolumn{1}{c}{+}  & -  & \multicolumn{1}{c}{-}  \\
   Li et al. \citeyearpar{24-Li}              	  & +  & \multicolumn{1}{c}{+}	& -	 & \multicolumn{1}{c}{-}  & -  & \multicolumn{1}{c}{-}  \\
  CommTrust \citeyearpar{26-Zhang}	      & +  & \multicolumn{1}{c}{+}	& -	 & \multicolumn{1}{c}{-}  & -  & \multicolumn{1}{c}{-}  \\
  Shabut et al. \citeyearpar{27-Shabut}	      & +  & \multicolumn{1}{c}{+}	& +  & \multicolumn{1}{c}{+}  & -  & \multicolumn{1}{c}{-}  \\
  \bottomrule
\end{tabular}
\end{minipage}
\end{table}

\textbf{RNP-based Trust Metrics.} Marti et al. \citeyearpar{15-Marti} proposed a voting reputation system wherein a node \(q\) would contact a set of nodes \(Res\) for their own local opinion on the responder \(r\), wherein the final reputation was calculated by summing each voter's rating in addition to the node \(q\)'s own feedback rating. J{\o}sang et al. \citeyearpar{55-Josang} defined the target \(Z\)'s reputation score at time \(t\) as \(R^t(Z)=(\delta+2a)/(\delta+\sigma+2)\), where \(\delta\) and \(\sigma\) denoted the numbers of positive and negative observations, and \(a\) expressed a prior or base rate to leverage the weight of positive rating. TrustWalker \cite{21-Jamali} estimated the rating for user \(u\) on target item \(i\) using different random walks, it summed the feedback information returned by different \(k\)-scoped random walks as the rating for the source user \(u\) on the target item \(i\). Wang et al. \citeyearpar{30-Wang} straightly defined the final trust of node \(i\) toward node \(j\) in \(d\) field as the integration of local trust value \(L_{ijd}\) and global trust value \(T_{jd}\) with a proportional factor. Liu et al. \citeyearpar{23-Liu} utilized the first-hop of trust transitivity \(T_{a1,a2}\) and the hop number to infer trust transitivity result \(T_{a1,a(j+1)}\) for a social trust path \(p_{(a_1, ..., a_{(j+1)})}\) as \(T_{a1,a(j+1)}\)= \(T_{a1,a2}\)+\(k_2\), where \(k_2\) denoted the slope of Deviation Line referring to the Base Line that started from coordination (1, \(T_{a1,a2}\)). Li et al. \citeyearpar{24-Li} utilized local trust degree (LTD) \(D_L(N_i, N_j)\) placed on node \(N_j\) from node \(N_i\) and feedback trust degree (FTD) \(R_U\)(\(N_i\), \(N_j\)) from the third-party nodes which had interactions with \(N_j\) to aggregate global trust. CommTrust \cite{26-Zhang} defined the overall trust score \(T\) for a seller as the weighted aggregation of dimension trust scores, i.e \(T = \sum\nolimits_{d = 1}^m {t_d \cdot w_d}\), where \(t_d\) and \(w_d\) represented the trust score and weight for the dimension \(d\) (\(d\)=1, ..., \(m\)). Then it utilized the beta function-based expectation to calculate trust score \(t_d  = (|\{ v_d  =  + 1\} | + m/2)/(n + m)\), where \(n\)=\(\{ v_d |v_d  =  + 1 \vee v_d  = -1\}|\) was the binary positive and negative ratings. Shabut et al. \citeyearpar{27-Shabut} adopted the direct and indirect trust to aggregate trust score \(T_{ij}\) for nodes \(i\) and \(j\), i.e. \(T_{ij}  = w_d  \cdot T_{ij}^d  + w_i  \cdot T_{ij}^i\), where \(w_d\)+\(w_i\)=1. The direct and indirect trust were inferred by the beta function-based expectation.
\begin{table}
\centering
\caption{Attack Analysis on CNP-based Trust Metrics}
\label{table:four}
\begin{minipage}{\columnwidth}
\renewcommand{\multirowsetup}{\centering}
\begin{tabular}{lclclclclclclcl}
  \toprule
  \multicolumn{1}{c|}{\multirow{2}{*}{Trust Metrics}} & \multicolumn{6}{|c}{Defense Countermeasure} \\
  \cline{2-7}
                   & \multicolumn{1}{|c|}{THR\_A}     & \multicolumn{1}{|c|}{THR\_B}  & \multicolumn{1}{|c|}{THR\_C}   & \multicolumn{1}{|c|}{THR\_D}     & \multicolumn{1}{|c|}{THR\_E}  & \multicolumn{1}{|c}{THR\_F}     \\
  \hline
  PeerTrustPSM \citeyearpar{38-Xiong}	& +	 & \multicolumn{1}{c}{+}  & -  & \multicolumn{1}{c}{+}  & -  & \multicolumn{1}{c}{-}  \\
  TrustGauard \citeyearpar{16-Srivatsa}	& +	 & \multicolumn{1}{c}{+}  & -  & \multicolumn{1}{c}{-}	& -	 & \multicolumn{1}{c}{-}  \\
  Yu et al. \citeyearpar{28-Yu}	            & +	 & \multicolumn{1}{c}{+}  & +  & \multicolumn{1}{c}{+}	& -	 &
  \multicolumn{1}{c}{-}  \\
  Li and Zhu \citeyearpar{36-Li}        & +	 & \multicolumn{1}{c}{+}  & +  & \multicolumn{1}{c}{+}	& -	 & \multicolumn{1}{c}{-}  \\
  DCMR \citeyearpar{89-Bao}     	    & +	 & \multicolumn{1}{c}{+}  & +  & \multicolumn{1}{c}{+}	& -	 & \multicolumn{1}{c}{-}  \\
  \bottomrule
\end{tabular}
\end{minipage}
\end{table}

\textbf{CNP-based Trust Metrics.} The personalized similarity was utilized to calculate feedback credibility of third-party recommenders in PeerTrustPSM \cite{38-Xiong}, TrustGauard \cite{16-Srivatsa} and Yu et al. \citeyearpar{28-Yu}. Li and Zhu \citeyearpar{36-Li} also utilized the cosine-based similarity as credibility for recommendation in body area networks. DCMR \cite{89-Bao} employed the similar users' similarity as the credibility to calculate the rating on an item in collaborative filtering.
\begin{table}
\centering
\caption{Attack Analysis on RUDP-based Trust Metrics}
\label{table:five}
\begin{minipage}{\columnwidth}
\renewcommand{\multirowsetup}{\centering}
\begin{tabular}{lclclclclclclcl}
  \toprule
  \multicolumn{1}{c|}{\multirow{2}{*}{Trust Metrics}} & \multicolumn{6}{|c}{Defense Countermeasure} \\
  \cline{2-7}
                   & \multicolumn{1}{|c|}{THR\_A}     & \multicolumn{1}{|c|}{THR\_B}  & \multicolumn{1}{|c|}{THR\_C}   & \multicolumn{1}{|c|}{THR\_D}     & \multicolumn{1}{|c|}{THR\_E}  & \multicolumn{1}{|c}{THR\_F}     \\
  \hline
  SocialTrust \citeyearpar{20-Caverlee}   & +  & \multicolumn{1}{c}{+}	& -	 & \multicolumn{1}{c}{-}  & -  & \multicolumn{1}{c}{-}  \\
  SORT \citeyearpar{25-Can}          	  & +  & \multicolumn{1}{c}{+}	& +  & \multicolumn{1}{c}{+}  & -  & \multicolumn{1}{c}{-}  \\
  GFTrust \citeyearpar{40-Jiang}	          & +  & \multicolumn{1}{c}{+}	& -	 & \multicolumn{1}{c}{-}  & -  & \multicolumn{1}{c}{-}  \\
  PageTrust \citeyearpar{57-Kerchove}	& +	 & \multicolumn{1}{c}{+}  & -  & \multicolumn{1}{c}{-}	& -	 & \multicolumn{1}{c}{-}  \\
  EigenTrust \citeyearpar{32-Kamvar}	& +	 & \multicolumn{1}{c}{+}  & -  & \multicolumn{1}{c}{-}	& -	 & \multicolumn{1}{c}{-}  \\
  PowerTrust \citeyearpar{19-Zhou}	    & +	 & \multicolumn{1}{c}{+}  & -  & \multicolumn{1}{c}{-}	& -	 & \multicolumn{1}{c}{-}  \\
  Guha et al. \citeyearpar{14-Guha}	        & +	 & \multicolumn{1}{c}{+}  & -  & \multicolumn{1}{c}{-}	& -	 & \multicolumn{1}{c}{-}  \\
  Dual-EigenRep \citeyearpar{22-Fan}	& +	 & \multicolumn{1}{c}{+}  & -  & \multicolumn{1}{c}{-}	& -  & \multicolumn{1}{c}{-}  \\
  Walter et al. \citeyearpar{87-Walter}	    & +	 & \multicolumn{1}{c}{+}  & -  & \multicolumn{1}{c}{-}	& -	 & \multicolumn{1}{c}{-}  \\
  \bottomrule
\end{tabular}
\end{minipage}
\end{table}

\textbf{RUDP-based Trust Metrics.} SocialTrust \cite{20-Caverlee} inferred a user \(i\)'s trust from the trust and relationship quality \(R(j)\) of other users, as well as the number of user \(j\)'s relationship, i.e. \(Tr_q (i) = \lambda \sum\limits_{j \in rel} {R(j) \cdot Tr_q (j)/|rel(j)| + (1 - \lambda )F(i)}\), where \(rel(j)\) denoted the set of contacts of user \(j\), \(F(i)\) represented the feedback rating aggregated by the trust group governing assessment. The relationship quality R(j) was a scoped random walk. SORT \cite{25-Can} defined participant \(p_i\)'s estimation about the reputation of \(p_j\) through collecting all recommendation trust from its acquaintance \(p_k\), namely \(er_{ij}  = \sum\nolimits_{p_k  \in T_i } {(rt_{ik}  \cdot \eta _{kj}  \cdot r_{kj} )}\), where \(\eta _{kj}\) was the number of \(p_k\)'s acquaintances which provided recommendations during the calculation of \(r_{kj}\), \(rt_{ik}\) denoted the recommendation trust from \(p_k\), and \(T_i\) was the set of trustworthy acquaintances selected by \(p_i\). GFTrust \cite{40-Jiang} mirrored trust propagation from nodes \(s\) to \(d\) as network flow with intermediate node \(v_i\) in the path \((s, v_1, ..., v_m, d)\), i.e. when a flow \(flw\) passed this path, the resulting flow would become \(flw \cdot \prod\nolimits_{i \in [1,m]} {(1 - leak(v_i ))}\), where \(leak(v_i)\) denoted the flow leakage function. PageRank \cite{132-Page} is the pioneering page ranking algorithm through propagating rank value from one page to its neighboring page(s) along the hyperlink or randomly to another non-hyperlinked page with a probability, it's a typically UDTP. On the basis of PageRank, PageTrust \cite{57-Kerchove} was proposed to infer the trust value for each page. EigenTrust \cite{32-Kamvar} adopted PageRank into trust management and accomplished UDTP through replacing the degree-based pairwise weight with the ratio of satisfied interactions. The similar manipulations also emerged in PowerTrust \cite{19-Zhou}, Dual-EigenRep \cite{22-Fan} and Guha et al. \citeyearpar{14-Guha}. Walter et al. \citeyearpar{87-Walter} calculated indirect trust via iterative computation of local trust matrix, the \(k\)th power of matrix represented the hops of UDTP.
\begin{table}
\centering
\caption{Attack Analysis on CUDP-based Trust Metrics}
\label{table:six}
\begin{minipage}{\columnwidth}
\renewcommand{\multirowsetup}{\centering}
\begin{tabular}{lclclclclclclcl}
  \toprule
  \multicolumn{1}{c|}{\multirow{2}{*}{Trust Metrics}} & \multicolumn{6}{|c}{Defense Countermeasure} \\
  \cline{2-7}
                   & \multicolumn{1}{|c|}{THR\_A}     & \multicolumn{1}{|c|}{THR\_B}  & \multicolumn{1}{|c|}{THR\_C}   & \multicolumn{1}{|c|}{THR\_D}     & \multicolumn{1}{|c|}{THR\_E}  & \multicolumn{1}{|c}{THR\_F}     \\
  \hline
  PeerTrustTVM \citeyearpar{38-Xiong} 	   & +  & \multicolumn{1}{c}{+}  & -  & \multicolumn{1}{c}{-}  & -  & \multicolumn{1}{c}{-}  \\
  Hu et al. \citeyearpar{12-hu}  	               & +  & \multicolumn{1}{c}{+}  & +  & \multicolumn{1}{c}{+}  & -	& \multicolumn{1}{c}{-}  \\
  EigenTrust\(^{++}\) \citeyearpar{34-fan} & +	& \multicolumn{1}{c}{+}	 & +  & \multicolumn{1}{c}{+}  & -	& \multicolumn{1}{c}{-}  \\
  ServiceTrust \citeyearpar{33-Su}	       & +	& \multicolumn{1}{c}{+}	 & +  & \multicolumn{1}{c}{+}  & -	& \multicolumn{1}{c}{-}  \\
  Deng et al. \citeyearpar{110-Deng}	           & +	& \multicolumn{1}{c}{+}  & +  & \multicolumn{1}{c}{+}  & -	& \multicolumn{1}{c}{-}  \\
  \bottomrule
\end{tabular}
\end{minipage}
\end{table}

\textbf{CUDP-based Trust Metrics.} PeerTrustTVM \cite{38-Xiong} employed each participant's trust ratio as feedback credibility to indicate how much trust it could propagate to its neighbor(s) using UDTP kernel. For a pair of participants, Hu et al. \citeyearpar{12-hu} defined feedback credibility by multiplying their transaction density factor and difference of ratings, then aggregated global trust through power-iteration-based matrix computation. EigenTrust\(^{++}\) \cite{34-fan} utilized the pairwise similarity as feedback credibility to weight raw direct trust for further global trust aggregation via UDTP kernel. ServiceTrust \cite{33-Su} defined positive similarity and negative similarity to merge the feedback credibility, subsequently aggregated global trust via UDTP kernel. Deng et al. \citeyearpar{110-Deng} associated the preference similarity to model trust degree for a pair of users. If no direct link existed, then the shortest path-based multiplication was adopted.
\begin{table}
\centering
\caption{Attack Analysis on RTCP-based Trust Metrics}
\label{table:seven}
\begin{minipage}{\columnwidth}
\renewcommand{\multirowsetup}{\centering}
\begin{tabular}{lclclclclclclcl}
  \toprule
  \multicolumn{1}{c|}{\multirow{2}{*}{Trust Metrics}} & \multicolumn{6}{|c}{Defense Countermeasure} \\
  \cline{2-7}
                   & \multicolumn{1}{|c|}{THR\_A}     & \multicolumn{1}{|c|}{THR\_B}  & \multicolumn{1}{|c|}{THR\_C}   & \multicolumn{1}{|c|}{THR\_D}     & \multicolumn{1}{|c|}{THR\_E}  & \multicolumn{1}{|c}{THR\_F}     \\
  \hline
  Chen et al. \citeyearpar{52-Chen}	            & +	 & \multicolumn{1}{c}{+}  & +  & \multicolumn{1}{c}{-}	& +	 &   \multicolumn{1}{c}{-}  \\
  Wang and Li \citeyearpar{90-Wang}    	 & +  & \multicolumn{1}{c}{+}  & -    & \multicolumn{1}{c}{-}    & -    & \multicolumn{1}{c}{-}  \\
  ReTrust \citeyearpar{29-He}	     & +  & \multicolumn{1}{c}{+}  & -    & \multicolumn{1}{c}{-}	 & -	& \multicolumn{1}{c}{-}  \\
  Chen et al. \citeyearpar{39-Chen}	     & +  & \multicolumn{1}{c}{+}  & +/-  & \multicolumn{1}{c}{+/-}  & +/-	&
  \multicolumn{1}{c}{+/-} \\
  AdaptiveTrust \citeyearpar{42-Chen} & +  & \multicolumn{1}{c}{+}  & +/-  & \multicolumn{1}{c}{+}    & +/-  &
  \multicolumn{1}{c}{+}  \\
  \bottomrule
\end{tabular}
\end{minipage}
\end{table}

\textbf{RTCP-based Trust Metrics.} Chen et al. \citeyearpar{52-Chen} proposed an inter-cluster recommendation trust concept and defined the total trust index from node \(N_i\) to \(N_j\) as \(\Gamma (N_i ,N_j)=\alpha T_D^{ij}+\beta T_R^j\), \(\alpha\), \(\beta\ge0\), \(\alpha+\beta\)=1, where \(\alpha\) and \(\beta\) were the impact weights of direct trust \(T_D^{ij}\) and recommendation trust \(T_R^j\) respectively. The inter-cluster recommendation trust for \(N_j\) was defined as \(T_R^j  = \sum\nolimits_{i = 1}^n {T_D^{hi} \cdot T_D^{ij} } /\sum\nolimits_{i = 1}^n {T_D^{hi} }\), where \(T_D^{hi}\)\(>\)\(H\), this condition indicated the cluster head (CH) would discard their recommendation to save bandwidth if the node's direct trust by CH was lower than a threshold value \(H\). Wang and Li \citeyearpar{90-Wang} calculated the aggregated rating by adopting Gaussian distribution-based upper control limit and lower control limit to filter out the marginal ratings out of the range of boundary. ReTrust \cite{29-He} utilized the similar range to block bad-mouthing recommendation, in addition to the indirect trust inference through trust propagation with the requirement that all direct trust between intermediate nodes must be greater than a threshold. Chen et al. \citeyearpar{39-Chen} used a threshold to select trustworthy recommenders to infer indirect trust, in addition to the consideration the trust between the originator node and recommender as a weight to multiply the recommendation trust. In AdaptiveTrust \cite {42-Chen}, the authors set a minimal honesty trust threshold (0.5) during time slot-based trust update.
\begin{table}
\centering
\caption{Attack Analysis on CTCP-based Trust Metrics}
\label{table:eight}
\begin{minipage}{\columnwidth}
\renewcommand{\multirowsetup}{\centering}
\begin{tabular}{lclclclclclclcl}
  \toprule
  \multicolumn{1}{c|}{\multirow{2}{*}{Trust Metrics}} & \multicolumn{6}{|c}{Defense Countermeasure} \\
  \cline{2-7}
                   & \multicolumn{1}{|c|}{THR\_A}     & \multicolumn{1}{|c|}{THR\_B}  & \multicolumn{1}{|c|}{THR\_C}   & \multicolumn{1}{|c|}{THR\_D}     & \multicolumn{1}{|c|}{THR\_E}  & \multicolumn{1}{|c}{THR\_F}     \\
  \hline
  Song et al. \citeyearpar{17-Song} 	        & +	 & \multicolumn{1}{c}{+}  & -  & \multicolumn{1}{c}{-}	& -	 & \multicolumn{1}{c}{-}  \\
  O'Donovan and Smyth \citeyearpar{92-O'Donovan}   	& +  & \multicolumn{1}{c}{+}  & +  & \multicolumn{1}{c}{+}  & -  & \multicolumn{1}{c}{-}  \\
  ServiveTrust\(^{++}\) \citeyearpar{43-Su}	& +	 & \multicolumn{1}{c}{+}  & +  & \multicolumn{1}{c}{+}	& -	 & \multicolumn{1}{c}{-}  \\
  GroupTrust \citeyearpar{4-Fan}    	    & +	 & \multicolumn{1}{c}{+}  & +  & \multicolumn{1}{c}{+}	& -	 & \multicolumn{1}{c}{-}  \\
  \bottomrule
\end{tabular}
\end{minipage}
\end{table}

\textbf{CTCP-based Trust Metrics.} Song et al. \citeyearpar{17-Song} utilized threshold-controlled selected local trust score \(t_{ji}\) placed on \(i\) from another participant \(j\) and corresponding aggregation weight to calculate the global trust, the weight was aggregated by the participant \(j\)'s trust, transaction date and amount. Consider participant \(j\)'s trust was under the computation simultaneously, therefore, the procedure of global trust in fact was a multiple iterations, which implied the trust was propagated through threshold-controlled edges. O'Donovan and Smyth \citeyearpar{92-O'Donovan} qualified which producer profiles were allowed to participate the rating recommendation process through predefining a threshold with which the item/profile-level trust values of candidate producer profiles compared, besides this, the qualified profiles also needed to take the harmonic mean of trust and similarity as recommendation credibility to infer the rating for an item in a consumer profile. ServiveTrust\(^{++}\) \cite{43-Su} employed the similarity as feedback credibility to weight the raw direct trust, in addition that a threshold over the holistic network was set to control the trust propagation. GroupTrust \cite{4-Fan} utilized the exponent-based feedback credibility, and proposed a fine-grained threshold-controlled trust propagation through studying the susceptible-infected-recovered model.
\section{DESIGN PRINCIPLES OF DECENTRALIZED TRUST MANAGEMENT} \label{sec:six}
Based on the formal analysis and experimental evaluation presented, we make three important observations and articulate the three design principles for effective trust management.

(1) Different threat modes have different adverse effects and different attack costs. Threat Models E and F have much more serious adverse effect than Threat Models A, B, C, D, because such attacks make it more difficult to differentiate mischievous participants from good participants in terms of both service provisioning or feedback rating behaviors. Thus, the robustness of decentralized trust management should be based on establishing trust by identifying both the list of priorities in terms of a range of services and the measurement for the service quality. In this paper we only cover the request serving and feedback rating as two types of services. We use the feedback rating to measure the request serving quality and we leverage feedback similarity as a way to measure feedback rating quality, which are vulnerable under the most strategically malicious Thread Models E and F.

(2) The feedback credibility-weighted local trust aggregation can effectively regulate the dishonest ratings when the fraction of malicious participants is much smaller than the fraction of good participants. However, the effectiveness of using the similarity based feedback credibility for regulating the local trust aggregation may no longer be effective when the fraction of good participants is out-numbered by the fraction of mischievous participants. This observation further indicates the importance of developing trust management algorithms that can tolerate unexpected errors and survive unexpected malicious attacks.

(3) The threshold controlled trust-propagation (TCTP) kernel provides the double-filtered function to regulate the trust propagation from good participants to the malicious participants. This enables the global trust scores of mischievous participants to be dramatically decreased, providing flexibility to control how trust is partially propagated over the network of participants through topological traversal. Thus, the TCTP kernel presents a more appropriate defense against those strategically mischievous attacks compared with the uniformly distributed trust propagation (UDTP) kernel.
\section{APPLICATION PROSPECT} \label{sec:seven}
\textbf{Edge Computing Trust.} One of the main attractions of edge computing is to improve the computation and processing cost and time by leveraging edge nodes instead of always connecting to the remote Cloud servers. Porambage et al. \citeyearpar{134-Porambage} pointed out "trust" as a significant mechanism in critical 5G use cases like remote surgeries, emergency autonomous vehicles, factory automation and tele-operated driving (e.g. drones). Several recent efforts have articulated the need to design appropriate trust management for edge cloud. Yan et al. \citeyearpar{108-Yan} articulated the role of trust management for reliable data fusion and mining, qualified services with context-awareness, and for enhancing user privacy and information security. Dang and Hoang \citeyearpar{109-Dang} demonstrated the use of trust management for data protection and performance improvement at edge servers. Garcia et al. \citeyearpar{111-Garcia} also presented the challenges for trust, security and privacy in edge-centric computing.

\textbf{Trust management in Blockchain Systems.} Blockchain technique \cite{85-Nakamoto} employs a decentralized P2P network to achieve consensus on a distributed public ledger of transactions through calculating the proofs of work for different peers (miners). The representative application of blockchain is the Bitcoin system. Users in the Bitcoin system are anonymous, and can use their public key hash as their pseudo-identity to interact with the system. However, the blockchain and the miner still confront some transactional risks at present, such as the double-spending problem and selfish mining problem \cite{140-Zhang, 86-Eyal, 71-Coleman, 112-Karame}.

One approach to mitigate such risks is to incorporate trust management into the peer to peer network of the blockchain system. For instance, consider the private blockchain and consortium blockchain scenarios, assume a service consumer \(u\) needs to pay a certain number of bitcoins to a service provider \(v\) using its address \(a_u\). If the transaction is successfully accomplished, i.e. user \(v\) receives bitcoins indeed from user \(u\), then user \(v\) will provide a positive rating to user \(u'\)s address \(a_u\), otherwise a negative rating if the payment does not accomplish successfully, i.e. user \(v\) does not receive bitcoins from user \(u\). Based on the feedback rating information, the blockchain system can produce a trust score for each anonymous address through employing our previously-introduced trust metrics, i.e. yield an overall estimation on the trustworthiness for each address. In subsequent transactions, service consumers can select the addresses with high trust scores as transacted targets, which can effectively block the transactions from the potentially-mischievous addresses with low trust scores, accordingly decline the risk of unsuccessful transactions.

The pull-in trust management can bring in two-facet advantages: i) keep the anonymity and un-traceability of users; and ii) guarantee the trustworthiness of transactional behaviors among users. The above is to rate the anonymous address, furthermore, we can also rate the user \(u\) with multiple addresses. Taking into account the linkability between addresses and users \cite{112-Karame, 113-Meiklejohn}, i.e. identify the ownership of different addresses for users, we can straightly accumulate the positive and negative ratings to multiple addresses offered by other users with which user \(u\) has had transactions, and aggregate a comprehensive trust score for user \(u\).

\textbf{Blockchain based Trust Management.\/} Trust scores and feedback ratings are important pieces of data that should be carefully protected in any trust management system. Blockchain technology can be utilized as an excellent mechanism to keep track of trust ratings and store them in the public and secure global ledger, such that a trust rating score once admitted into the blockchain, it will be absolutely secure from malicious modification and compromises. We argue that this is an interesting research and development project with high practical relevance.
\section{CONCLUSIONS AND FUTURE WORK} \label{sec:eight}
We describe decentralized trust management models and their efficiency and robustness from three unique perspectives. First, we study the risk factors and adverse effects of six common threat models. Second, we review the representative trust aggregation models and trust metrics. Third, we present an in-depth analysis and comparison of these reference trust aggregation methods with respect to effectiveness and robustness. We argue that our comparative study advances the understanding of adverse effects of present and future threats and the robustness of different trust metrics. It may also serve as a guideline for research and development of next generation trust aggregation algorithms and services in the anticipation of risk factors and mischievous threats.

\begin{acks}
The authors would like to thank Prof. Dr. Fang and anonymous reviewers for their helpful suggestions and comments that significantly improved the presentation of the paper. The authors from Chinese Academy of Sciences are supported by the National Natural Science Foundation of China under Grant No.: 61702470, 61472403. The author from Georgia Institute of Technology, USA is partially funded by the USA National Natural Science Foundation under Grants 1547102, SaTC 1564097 and IBM faculty award.
\end{acks}

\bibliographystyle{ACM-Reference-Format}
\bibliography{sample-base}





\end{document}